 \documentclass[aps,prd,twocolumn,amsmath,amssymb,amsfonts,nofootinbib,superscriptaddress]{revtex4-1}
\usepackage{float}
\usepackage{comment}

\usepackage{graphicx}
\usepackage{dcolumn}
\usepackage{bm}
\usepackage{url}
\usepackage{subfigure}
\usepackage{float}
\usepackage{amssymb}
\usepackage{amsmath}
\usepackage{longtable}
\usepackage{rotating}
\usepackage{color}
\usepackage{xcolor}
\usepackage{epsfig}
\usepackage{epsf}
\usepackage{fancyhdr}
\usepackage{hyperref}

\def\be{\begin{equation}}
\def\bea{\begin{eqnarray}}
\def\eea{\end{eqnarray}}
\def\ee{\end{equation}}
\def\bi{\begin{itemize}}
\def\ei{\end{itemize}}

\def\bn{\begin{enumerate}}
\def\en{\end{enumerate}}

\def\be{\begin{equation}}
\def\ee{\end{equation}}
\def\bea{\begin{eqnarray}}
\def\eea{\end{eqnarray}}

\def\beq{\begin{equation}}
\def\eeq{\end{equation}}

\begin{document}

\title{First narrow-band search for continuous gravitational waves from known pulsars in advanced detector data}

\begin{abstract} 
Spinning neutron stars asymmetric with respect to their rotation axis are potential sources of continuous gravitational waves for ground-based interferometric detectors. In the case of known pulsars a fully coherent search, based on matched filtering, which uses the 
position and rotational parameters obtained from electromagnetic observations, can be carried out. Matched filtering maximizes the signal-to-noise (SNR) ratio, but a large sensitivity loss is expected in case of even a very small mismatch between the assumed and the true signal parameters. For this reason, {\it narrow-band} analyses methods have been developed, allowing a fully coherent search for gravitational waves from known pulsars over a fraction of a hertz and several spin-down values. In this paper we describe a narrow-band search of eleven pulsars using data from Advanced LIGO's first observing run. Although we have found several initial outliers, further studies show no significant evidence for the presence of a gravitational wave signal. Finally, we have placed upper limits on the signal strain amplitude lower than the spin-down limit for 5 of the 11 targets over the bands searched: in the case of J1813-1749 the spin-down limit has been beaten for the first time. For an additional 3 targets, the median upper limit across the search bands is below the spin-down limit. This is the most sensitive narrow-band search for continuous gravitational waves carried out so far.
\end{abstract}


\author{%
B.~P.~Abbott,$^{1}$  
R.~Abbott,$^{1}$  
T.~D.~Abbott,$^{2}$  
F.~Acernese,$^{3,4}$ 
K.~Ackley,$^{5,6}$  
C.~Adams,$^{7}$  
T.~Adams,$^{8}$ 
P.~Addesso,$^{9}$  
R.~X.~Adhikari,$^{1}$  
V.~B.~Adya,$^{10}$
% ...
\\
\medskip
(LIGO Scientific Collaboration and Virgo Collaboration) 
\\
\medskip
{${}^{\dag}$Deceased, February 2017. }%
{${}^{\ddag}$Deceased, December 2016. }%
}\noaffiliation
\affiliation {LIGO, California Institute of Technology, Pasadena, CA 91125, USA }
\affiliation {Louisiana State University, Baton Rouge, LA 70803, USA }
\affiliation {Universit\`a di Salerno, Fisciano, I-84084 Salerno, Italy }
\affiliation {INFN, Sezione di Napoli, Complesso Universitario di Monte S.Angelo, I-80126 Napoli, Italy }
\affiliation {University of Florida, Gainesville, FL 32611, USA }
\affiliation {OzGrav, School of Physics \& Astronomy, Monash University, Clayton 3800, Victoria, Australia }
\affiliation {LIGO Livingston Observatory, Livingston, LA 70754, USA }
\affiliation {Laboratoire d'Annecy-le-Vieux de Physique des Particules (LAPP), Universit\'e Savoie Mont Blanc, CNRS/IN2P3, F-74941 Annecy, France }
\affiliation {University of Sannio at Benevento, I-82100 Benevento, Italy and INFN, Sezione di Napoli, I-80100 Napoli, Italy }
\affiliation {Max Planck Institute for Gravitational Physics (Albert Einstein Institute), D-30167 Hannover, Germany }
\affiliation {The University of Mississippi, University, MS 38677, USA }
\affiliation {NCSA, University of Illinois at Urbana-Champaign, Urbana, IL 61801, USA }
\affiliation {University of Cambridge, Cambridge CB2 1TN, United Kingdom }
\affiliation {Nikhef, Science Park, 1098 XG Amsterdam, The Netherlands }
\affiliation {LIGO, Massachusetts Institute of Technology, Cambridge, MA 02139, USA }
\affiliation {Instituto Nacional de Pesquisas Espaciais, 12227-010 S\~{a}o Jos\'{e} dos Campos, S\~{a}o Paulo, Brazil }
\affiliation {Gran Sasso Science Institute (GSSI), I-67100 L'Aquila, Italy }
\affiliation {INFN, Laboratori Nazionali del Gran Sasso, I-67100 Assergi, Italy }
\affiliation {Inter-University Centre for Astronomy and Astrophysics, Pune 411007, India }
\affiliation {University of Wisconsin-Milwaukee, Milwaukee, WI 53201, USA }
\affiliation {Leibniz Universit\"at Hannover, D-30167 Hannover, Germany }
\affiliation {Universit\`a di Pisa, I-56127 Pisa, Italy }
\affiliation {INFN, Sezione di Pisa, I-56127 Pisa, Italy }
\affiliation {OzGrav, Australian National University, Canberra, Australian Capital Territory 0200, Australia }
\affiliation {Laboratoire des Mat\'eriaux Avanc\'es (LMA), CNRS/IN2P3, F-69622 Villeurbanne, France }
\affiliation {SUPA, University of the West of Scotland, Paisley PA1 2BE, United Kingdom }
\affiliation {LAL, Univ. Paris-Sud, CNRS/IN2P3, Universit\'e Paris-Saclay, F-91898 Orsay, France }
\affiliation {California State University Fullerton, Fullerton, CA 92831, USA }
\affiliation {European Gravitational Observatory (EGO), I-56021 Cascina, Pisa, Italy }
\affiliation {Chennai Mathematical Institute, Chennai 603103, India }
\affiliation {Universit\`a di Roma Tor Vergata, I-00133 Roma, Italy }
\affiliation {INFN, Sezione di Roma Tor Vergata, I-00133 Roma, Italy }
\affiliation {Universit\"at Hamburg, D-22761 Hamburg, Germany }
\affiliation {INFN, Sezione di Roma, I-00185 Roma, Italy }
\affiliation {Cardiff University, Cardiff CF24 3AA, United Kingdom }
\affiliation {Embry-Riddle Aeronautical University, Prescott, AZ 86301, USA }
\affiliation {Max Planck Institute for Gravitational Physics (Albert Einstein Institute), D-14476 Potsdam-Golm, Germany }
\affiliation {APC, AstroParticule et Cosmologie, Universit\'e Paris Diderot, CNRS/IN2P3, CEA/Irfu, Observatoire de Paris, Sorbonne Paris Cit\'e, F-75205 Paris Cedex 13, France }
\affiliation {Korea Institute of Science and Technology Information, Daejeon 34141, Korea }
\affiliation {West Virginia University, Morgantown, WV 26506, USA }
\affiliation {Universit\`a di Perugia, I-06123 Perugia, Italy }
\affiliation {INFN, Sezione di Perugia, I-06123 Perugia, Italy }
\affiliation {Syracuse University, Syracuse, NY 13244, USA }
\affiliation {University of Minnesota, Minneapolis, MN 55455, USA }
\affiliation {SUPA, University of Glasgow, Glasgow G12 8QQ, United Kingdom }
\affiliation {LIGO Hanford Observatory, Richland, WA 99352, USA }
\affiliation {Caltech CaRT, Pasadena, CA 91125, USA }
\affiliation {Wigner RCP, RMKI, H-1121 Budapest, Konkoly Thege Mikl\'os \'ut 29-33, Hungary }
\affiliation {Columbia University, New York, NY 10027, USA }
\affiliation {Stanford University, Stanford, CA 94305, USA }
\affiliation {Universit\`a di Camerino, Dipartimento di Fisica, I-62032 Camerino, Italy }
\affiliation {Universit\`a di Padova, Dipartimento di Fisica e Astronomia, I-35131 Padova, Italy }
\affiliation {INFN, Sezione di Padova, I-35131 Padova, Italy }
\affiliation {Institute of Physics, E\"otv\"os University, P\'azm\'any P. s. 1/A, Budapest 1117, Hungary }
\affiliation {Nicolaus Copernicus Astronomical Center, Polish Academy of Sciences, 00-716, Warsaw, Poland }
\affiliation {Rochester Institute of Technology, Rochester, NY 14623, USA }
\affiliation {University of Birmingham, Birmingham B15 2TT, United Kingdom }
\affiliation {INFN, Sezione di Genova, I-16146 Genova, Italy }
\affiliation {RRCAT, Indore MP 452013, India }
\affiliation {Faculty of Physics, Lomonosov Moscow State University, Moscow 119991, Russia }
\affiliation {SUPA, University of Strathclyde, Glasgow G1 1XQ, United Kingdom }
\affiliation {The Pennsylvania State University, University Park, PA 16802, USA }
\affiliation {OzGrav, University of Western Australia, Crawley, Western Australia 6009, Australia }
\affiliation {Department of Astrophysics/IMAPP, Radboud University Nijmegen, P.O. Box 9010, 6500 GL Nijmegen, The Netherlands }
\affiliation {Artemis, Universit\'e C\^ote d'Azur, Observatoire C\^ote d'Azur, CNRS, CS 34229, F-06304 Nice Cedex 4, France }
\affiliation {Institut FOTON, CNRS, Universit\'e de Rennes 1, F-35042 Rennes, France }
\affiliation {Washington State University, Pullman, WA 99164, USA }
\affiliation {University of Oregon, Eugene, OR 97403, USA }
\affiliation {Laboratoire Kastler Brossel, UPMC-Sorbonne Universit\'es, CNRS, ENS-PSL Research University, Coll\`ege de France, F-75005 Paris, France }
\affiliation {Carleton College, Northfield, MN 55057, USA }
\affiliation {OzGrav, University of Adelaide, Adelaide, South Australia 5005, Australia }
\affiliation {Astronomical Observatory Warsaw University, 00-478 Warsaw, Poland }
\affiliation {VU University Amsterdam, 1081 HV Amsterdam, The Netherlands }
\affiliation {University of Maryland, College Park, MD 20742, USA }
\affiliation {Center for Relativistic Astrophysics, Georgia Institute of Technology, Atlanta, GA 30332, USA }
\affiliation {Universit\'e Claude Bernard Lyon 1, F-69622 Villeurbanne, France }
\affiliation {Universit\`a di Napoli `Federico II,' Complesso Universitario di Monte S.Angelo, I-80126 Napoli, Italy }
\affiliation {NASA Goddard Space Flight Center, Greenbelt, MD 20771, USA }
\affiliation {RESCEU, University of Tokyo, Tokyo, 113-0033, Japan. }
\affiliation {Tsinghua University, Beijing 100084, China }
\affiliation {Texas Tech University, Lubbock, TX 79409, USA }
\affiliation {Kenyon College, Gambier, OH 43022, USA }
\affiliation {Departamento de Astronom\'{\i }a y Astrof\'{\i }sica, Universitat de Val\`encia, E-46100 Burjassot, Val\`encia, Spain }
\affiliation {Museo Storico della Fisica e Centro Studi e Ricerche Enrico Fermi, I-00184 Roma, Italy }
\affiliation {National Tsing Hua University, Hsinchu City, 30013 Taiwan, Republic of China }
\affiliation {Charles Sturt University, Wagga Wagga, New South Wales 2678, Australia }
\affiliation {Center for Interdisciplinary Exploration \& Research in Astrophysics (CIERA), Northwestern University, Evanston, IL 60208, USA }
\affiliation {University of Chicago, Chicago, IL 60637, USA }
\affiliation {Pusan National University, Busan 46241, Korea }
\affiliation {The Chinese University of Hong Kong, Shatin, NT, Hong Kong }
\affiliation {INAF, Osservatorio Astronomico di Padova, I-35122 Padova, Italy }
\affiliation {INFN, Trento Institute for Fundamental Physics and Applications, I-38123 Povo, Trento, Italy }
\affiliation {Dipartimento di Fisica, Universit\`a degli Studi di Genova, I-16146 Genova, Italy }
\affiliation {OzGrav, University of Melbourne, Parkville, Victoria 3010, Australia }
\affiliation {Universit\`a di Roma `La Sapienza,' I-00185 Roma, Italy }
\affiliation {Universit\'e Libre de Bruxelles, Brussels 1050, Belgium }
\affiliation {Sonoma State University, Rohnert Park, CA 94928, USA }
\affiliation {Departamento de Matem\'aticas, Universitat de Val\`encia, E-46100 Burjassot, Val\`encia, Spain }
\affiliation {Montana State University, Bozeman, MT 59717, USA }
\affiliation {Universitat de les Illes Balears, IAC3---IEEC, E-07122 Palma de Mallorca, Spain }
\affiliation {The University of Texas Rio Grande Valley, Brownsville, TX 78520, USA }
\affiliation {Bellevue College, Bellevue, WA 98007, USA }
\affiliation {Institute for Plasma Research, Bhat, Gandhinagar 382428, India }
\affiliation {The University of Sheffield, Sheffield S10 2TN, United Kingdom }
\affiliation {Dipartimento di Scienze Matematiche, Fisiche e Informatiche, Universit\`a di Parma, I-43124 Parma, Italy }
\affiliation {INFN, Sezione di Milano Bicocca, Gruppo Collegato di Parma, I-43124 Parma, Italy }
\affiliation {California State University, Los Angeles, 5151 State University Dr, Los Angeles, CA 90032, USA }
\affiliation {Universit\`a di Trento, Dipartimento di Fisica, I-38123 Povo, Trento, Italy }
\affiliation {Montclair State University, Montclair, NJ 07043, USA }
\affiliation {National Astronomical Observatory of Japan, 2-21-1 Osawa, Mitaka, Tokyo 181-8588, Japan }
\affiliation {Canadian Institute for Theoretical Astrophysics, University of Toronto, Toronto, Ontario M5S 3H8, Canada }
\affiliation {Observatori Astron\`omic, Universitat de Val\`encia, E-46980 Paterna, Val\`encia, Spain }
\affiliation {School of Mathematics, University of Edinburgh, Edinburgh EH9 3FD, United Kingdom }
\affiliation {University and Institute of Advanced Research, Koba Institutional Area, Gandhinagar Gujarat 382007, India }
\affiliation {IISER-TVM, CET Campus, Trivandrum Kerala 695016, India }
\affiliation {University of Szeged, D\'om t\'er 9, Szeged 6720, Hungary }
\affiliation {International Centre for Theoretical Sciences, Tata Institute of Fundamental Research, Bengaluru 560089, India }
\affiliation {University of Michigan, Ann Arbor, MI 48109, USA }
\affiliation {Tata Institute of Fundamental Research, Mumbai 400005, India }
\affiliation {INAF, Osservatorio Astronomico di Capodimonte, I-80131, Napoli, Italy }
\affiliation {Universit\`a degli Studi di Urbino `Carlo Bo,' I-61029 Urbino, Italy }
\affiliation {INFN, Sezione di Firenze, I-50019 Sesto Fiorentino, Firenze, Italy }
\affiliation {Physik-Institut, University of Zurich, Winterthurerstrasse 190, 8057 Zurich, Switzerland }
\affiliation {American University, Washington, D.C. 20016, USA }
\affiliation {University of Bia{\l }ystok, 15-424 Bia{\l }ystok, Poland }
\affiliation {University of Southampton, Southampton SO17 1BJ, United Kingdom }
\affiliation {University of Washington Bothell, 18115 Campus Way NE, Bothell, WA 98011, USA }
\affiliation {Institute of Applied Physics, Nizhny Novgorod, 603950, Russia }
\affiliation {Korea Astronomy and Space Science Institute, Daejeon 34055, Korea }
\affiliation {Inje University Gimhae, South Gyeongsang 50834, Korea }
\affiliation {National Institute for Mathematical Sciences, Daejeon 34047, Korea }
\affiliation {NCBJ, 05-400 \'Swierk-Otwock, Poland }
\affiliation {Institute of Mathematics, Polish Academy of Sciences, 00656 Warsaw, Poland }
\affiliation {Hillsdale College, Hillsdale, MI 49242, USA }
\affiliation {Hanyang University, Seoul 04763, Korea }
\affiliation {Seoul National University, Seoul 08826, Korea }
\affiliation {NASA Marshall Space Flight Center, Huntsville, AL 35811, USA }
\affiliation {ESPCI, CNRS, F-75005 Paris, France }
\affiliation {Southern University and A\&M College, Baton Rouge, LA 70813, USA }
\affiliation {College of William and Mary, Williamsburg, VA 23187, USA }
\affiliation {Centre Scientifique de Monaco, 8 quai Antoine Ier, MC-98000, Monaco }
\affiliation {Indian Institute of Technology Madras, Chennai 600036, India }
\affiliation {IISER-Kolkata, Mohanpur, West Bengal 741252, India }
\affiliation {Whitman College, 345 Boyer Avenue, Walla Walla, WA 99362 USA }
\affiliation {Indian Institute of Technology Bombay, Powai, Mumbai, Maharashtra 400076, India }
\affiliation {Scuola Normale Superiore, Piazza dei Cavalieri 7, I-56126 Pisa, Italy }
\affiliation {Universit\'e de Lyon, F-69361 Lyon, France }
\affiliation {Hobart and William Smith Colleges, Geneva, NY 14456, USA }
\affiliation {OzGrav, Swinburne University of Technology, Hawthorn VIC 3122, Australia }
\affiliation {Janusz Gil Institute of Astronomy, University of Zielona G\'ora, 65-265 Zielona G\'ora, Poland }
\affiliation {University of Washington, Seattle, WA 98195, USA }
\affiliation {King's College London, University of London, London WC2R 2LS, United Kingdom }
\affiliation {Indian Institute of Technology, Gandhinagar Ahmedabad Gujarat 382424, India }
\affiliation {Indian Institute of Technology Hyderabad, Sangareddy, Khandi, Telangana 502285, India }
\affiliation {International Institute of Physics, Universidade Federal do Rio Grande do Norte, Natal RN 59078-970, Brazil }
\affiliation {Andrews University, Berrien Springs, MI 49104, USA }
\affiliation {Universit\`a di Siena, I-53100 Siena, Italy }
\affiliation {Trinity University, San Antonio, TX 78212, USA }
\affiliation {Abilene Christian University, Abilene, TX 79699, USA }
\affiliation {Colorado State University, Fort Collins, CO 80523, USA }

\maketitle

\section{Introduction}

\begin{figure*}[!hbpt]
\includegraphics[width=0.7\textwidth]{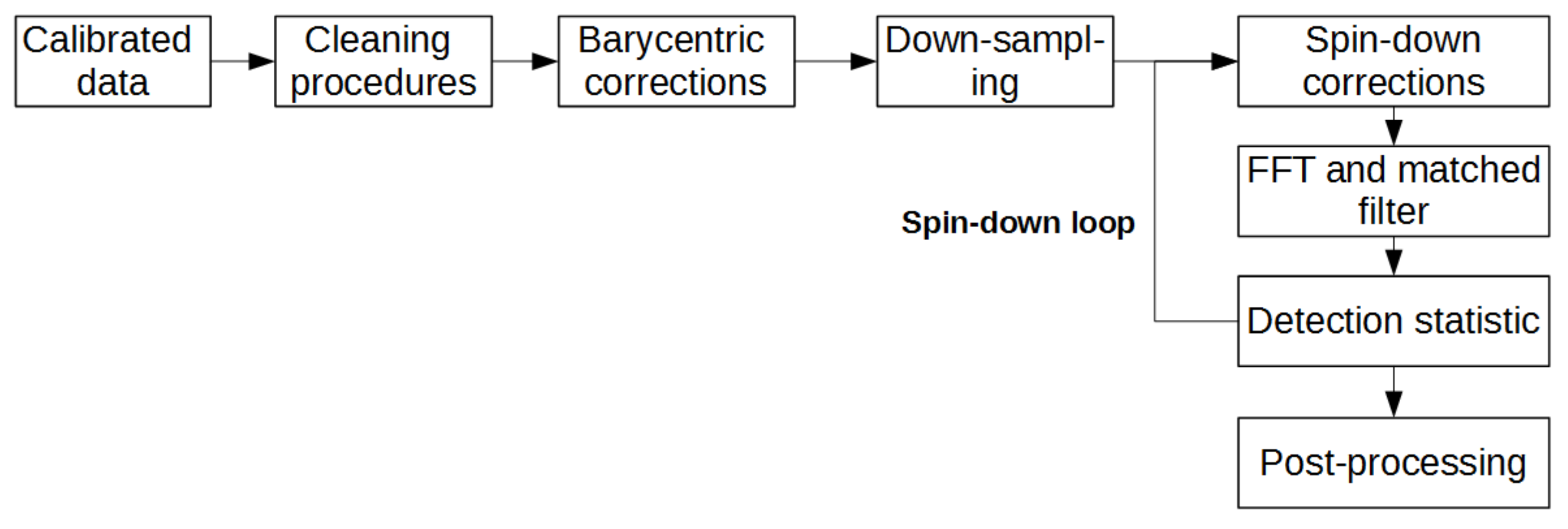}
\caption{Simplified flowchart of the narrow-band search pipeline for CW. The method relies on the use of FFTs to simultaneously compute the detection statistic, for each given spin-down value, over the full explored frequency range. See \cite{mastrogiovanni:method} for more details on the method.}
\label{fig:flow}
\end{figure*}

On  September 14th 2015 the gravitational wave (GW) signal emitted by a binary black hole merger was detected by the LIGO interferometers (IFOs) \cite{abbot:detection} followed on 26th December 2015, by the detection of a second event again associated to a binary black hole merger\cite{abbot:detection2}, thus opening the era of gravitational waves astronomy. More recently, the detection of a third binary black hole merger on Jan 4th 2017 has been announced\cite{ref:GW170104}.
Binary black hole mergers, however, are not the only detectable sources of GW. Among the potential sources of GW there are also spinning neutron stars (NS) asymmetric with respect to their rotation axis. These sources are expected to emit nearly monochromatic continuous waves (CW), with a frequency at a given fixed ratio with respect to the star's rotational frequency, e.g. two times the rotational frequency for an asymmetric NS rotating around one of its principal axis of inertia. Different flavors of CW searches exist, depending on the degree of knowledge on the source parameters. {\it Targeted} searches assume source position and rotational parameters to be known with high accuracy, while {\it all-sky} searches aim at neutron stars with no observed electromagnetic counterpart. Various intermediate searches have also been developed. Among these, \textit{narrow-band} searches are an extension of targeted searches for which the position of the source is accurately known but, the rotational parameters are slightly uncertain. Narrow-band searches allow  for a possible small mismatch between the GW rotational parameters and those inferred from electromagnetic observations. This can be crucial if, for instance, the CW signal is emitted by a freely precessing neutron star \cite{LVC:old}, or in the case no updated ephemeris is available for a given pulsar. In both cases a targeted search could assume wrong rotational parameters, resulting in a significant sensitivity loss. In this paper we present the results of a fully coherent, narrow-band search for 11 known pulsars using data from the first observation run (O1) of the Advanced LIGO detectors\cite{LIGO:adv}. The paper is organized as follows. In Sec. \ref{sec:signal} we briefly summarise the main concepts of the analysis. Sec. \ref{sec:method} is dedicated to an outline of the analysis method. Sec. \ref{sec:targets} describes the selected pulsars.
In Sec. \ref{sec:results} we discuss the analysis results, while the reader can refer to the Appendix for some technical details on the computation of upper limits. Finally, Sec. \ref{sec:conclusion} is dedicated to the conclusions and future prospects. 

\section{Background}
\label{sec:signal}
The GW signal emitted by an asymmetric  spinning NS can be written, following the formalism first introduced in \cite{pia:articolo}, as the real part of:
\begin{equation}
h(t)= H_0 ( H^+ A_+ (t) + H^\times A_\times (t)) e^{2 \pi i f_{\mathrm{gw}} (t) t+i \phi_0}
\label{eq:Hgrande}
\end{equation}
where $f_\mathrm{gw} (t)$ is the GW frequency, $\phi_0$ an initial phase. The polarisation amplitudes $H^+, H^\times$ are given by:
\begin{eqnarray*}
H^+ =\frac{\cos(2 \psi) - i \eta \sin (2 \psi)}{\sqrt{1+\eta^2}},  \quad  H^\times =\frac{\sin(2 \psi) - i \eta \cos (2 \psi)}{\sqrt{1+\eta^2}},  
\end{eqnarray*}
$\eta$ being the ratio of the polarisation ellipse semi-minor to semi-major axis and $\psi$ the polarization angle, defined as the direction of the major axis with respect to the celestial parallel of the source (measured counter-clockwise). The detector \textit{sidereal response} to the GW polarisations is encoded in the functions $A_+ (t), A_\times (t)$. It can be shown that the waveform defined by Eq. \ref{eq:Hgrande} is equivalent to the GW signal expressed in the more standard formalism of \cite{matt:O1_known}, given the following relations: 
\begin{equation}
\eta=-\frac{2\cos \iota}{1+\cos^2 \iota},
\label{eq:etaiota}
\end{equation}
where $\iota$ is the angle between the line of sight and the star rotation axis, and
\begin{equation}
H_0=h_0 \sqrt{ \frac{1+6 \cos ^2 \iota + \cos^4 \iota}{4}} 
\end{equation}
 with
\begin{equation}
h_0=\frac{1}{d} \frac{4 \pi^2 G }{c^4} I_\mathrm{zz} f_\mathrm{gw}^2 \epsilon,
\label{eq:GW_amplitude}
\end{equation}
where $d,I_\mathrm{zz}$ and $\epsilon$ are respectively the star's distance, its moment of inertia with respect to the rotation axis and the {\it ellipticity}, which measures the star's degree of asymmetry.  The signal at the detector is not monochromatic, i.e. the frequency $f_\mathrm{gw} (t)$ in Eq. \ref{eq:Hgrande} is a function of time. In fact the signal is modulated by several effects, such as the \textit{R\"{o}mer delay} due to the detector motion and the source's intrinsic spin-down due to the rotational energy loss from the source. In order to recover all the signal to noise ratio all these effects must be properly taken into account. If we have a measure of the pulsar rotational frequency $f_\mathrm{rot}$, frequency derivative $\dot{f}_\mathrm{rot}$ and distance $d$, the GW signal amplitude can be constrained, assuming that all the rotational energy is lost via gravitational radiation. This strict upper limit, called {\it spin-down limit}, is given by\cite{krolak:fstat}: 
\begin{equation}
h_{sd}=8.06 \cdot 10^{-19} I_{38}^{1/2} \bigg[\frac{1\mathrm{kpc}}{d} \bigg] \bigg[\frac{\dot{f}_\mathrm{rot}}{\mathrm{Hz/s}} \bigg]^{1/2} \bigg[\frac{\mathrm{Hz}}{f_\mathrm{rot}} \bigg]^{1/2}
\label{eq:sd_limit}
\end{equation} 
where $I_{38}$ is the star moment of inertia in unit of $10^{38} \mathrm{kg \, m^2}$. The corresponding spin-down limit on the star equatorial fiducial ellipticity can be easily obtained from Eq. \ref{eq:GW_amplitude}.
\begin{equation}
\epsilon_{sd}=0.237 \, I_{38}^{-1} \bigg[ \frac{h_\mathrm{sd}}{10^{-24}}\bigg]  \bigg[\frac{\mathrm{Hz}}{f_\mathrm{rot}} \bigg]^{2}  \bigg[\frac{d}{1 \mathrm{kpc}} \bigg] \, _.
\end{equation} 
Even in the absence of a detection, establishing an amplitude upper limit below the spin-down limit for a given source is an important milestone, as it allows us to put a non-trivial constraint on the fraction of rotational energy lost through GWs.

\section{The analysis}
\label{sec:method}
\begin{figure*}[!htpb]
\centering
\includegraphics[scale=0.35]{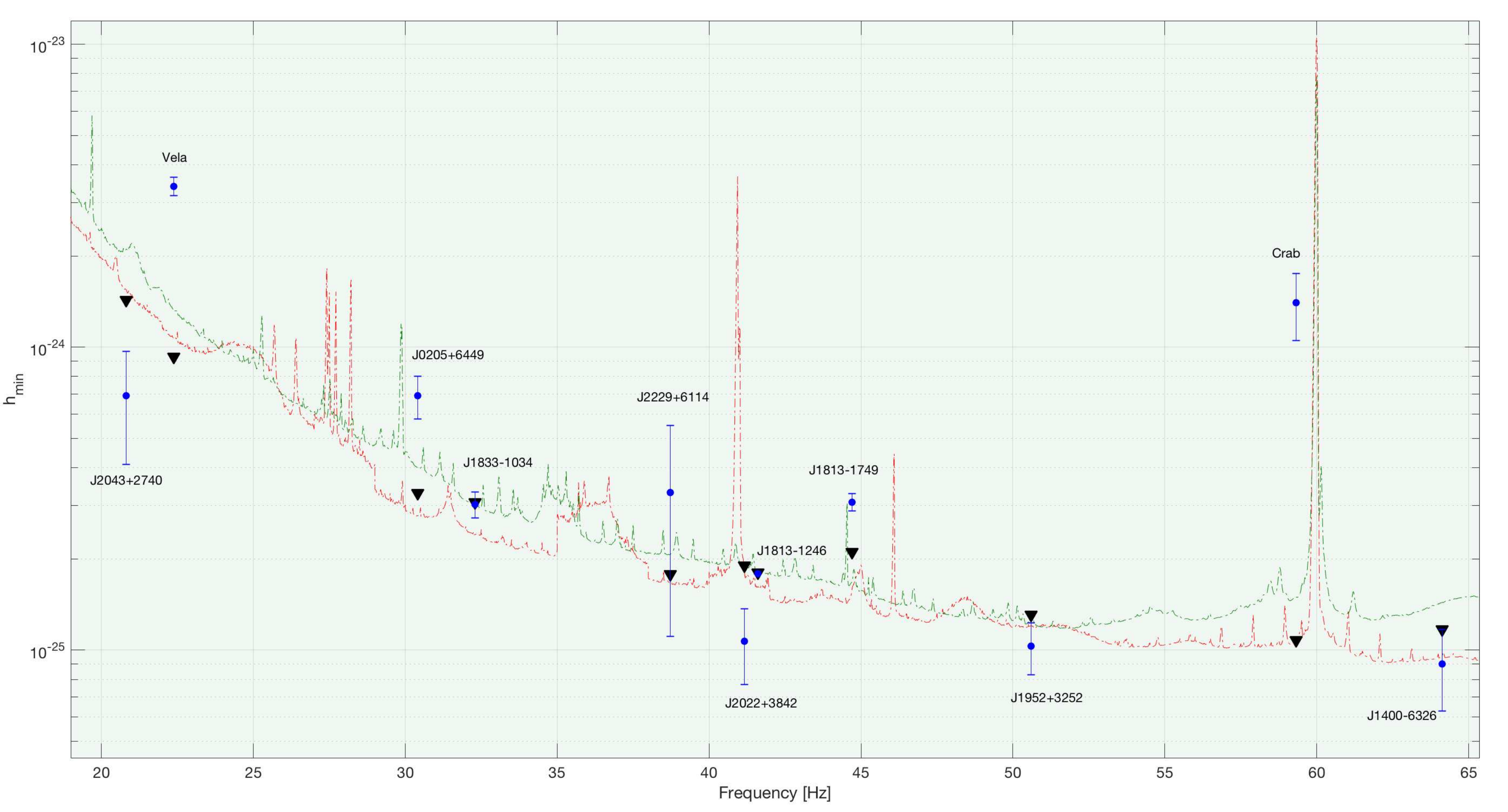}
\caption{Blue points: Value of the theoretical spin-down limit computed for the 11 known pulsars in our analysis, corresponding to Tab. \ref{tab:tab1}, error bars correspond to 1$\sigma$ confidence level. Black triangles: median over the analysed frequency band of the upper-limits on the GW amplitude, corresponding to Tab. \ref{tab:tab3}. Red dashed line: Estimated sensitivity at 95\% confidence level of a narrow-band search using data from LIGO H. Green dashed line: Estimated sensitivity at 95\% confidence level of a narrow-band search using data from LIGO L.}
\label{fig:spin_down_compar}
\end{figure*}
The results discussed in this paper have been obtained by searching for CW signals from 11 known pulsars using data from the O1 run from the Advanced LIGO detectors (Hanford - LIGO H, and Livingston - LIGO L jointly). The run started on September 12th 2015 at 01:25:03 UTC and 18:29:03 UTC, respectively, and finished on  January 19th 2016 at 17:07:59. LIGO H had a duty cycle of $\sim$60\% and LIGO L had a duty cycle of $\sim$51\%, which correspond respectively to 72 and 62 days of science data available for the analysis. In this paper we have used an initial calibration of the data \cite{abbott:calibration}. In order to perform joint search between the two detectors a common period from September 13th 2015  to  January 12th 2016 \footnote{An exception is pulsar J0205+6449, see later.}, with a total observation time of about $T_\mathrm{obs} \approx 121 \, \text{days}$ is selected. The natural frequency and spin-down grid spacings of the search are $\delta f=1/T_\mathrm{obs} \approx 9.5 \cdot 10^{-8} \, \text{Hz}$ and $\delta \dot{f} =1/T_\mathrm{obs}^2\approx 4.57 \cdot 10^{-15} \, \text{Hz/s}$. A follow-up analysis based on the LIGO's second observation Run (O2) has been carried out, for this dataset we have analysed data from December 16th 2016 to May 8th 2017, more details will be given in Appendix C.
The analysis pipeline consists of several steps, schematically depicted in Fig. \ref{fig:flow}, which we summarise here. The starting point is a collection of FFTs obtained from several interlaced data chunks (the short FFT Database - SFDB) built from calibrated detector data chunks of duration 1024 seconds \cite{astone:short}. At this stage, a first cleaning procedure is applied to the data in order to remove large, short-duration disturbances, that could reduce the search sensitivity. A frequency band is then extracted from the SFDBs covering typically a range larger (of the order of a factor of 2) than the frequency region analysed in the narrow-band search. The actual search frequency and spin-down bands, $\Delta f $ and $\Delta \dot{f}$, around the reference values, $f_0 $ and $ \dot{f}_0$, have been chosen according to the following relations \cite{rob:obs}:
 \begin{eqnarray}
&  \Delta f= 2 f_0 \delta \\
& \Delta \dot{f}= 2 \dot{f}_0 \delta, 
\end{eqnarray}
$\delta$ being a factor parametrizing a possible discrepancy between the GW rotation parameters and those derived from electromagnetic observations. Previous narrowband searches used values of $\delta$ of the order $\sim \mathcal{O}(10^{-4})$, motivated partly by astrophysical considerations\cite{LVC:old}, and partly by computational limitations \cite{rob:method}.  Here we exploit the high computation efficiency of our pipeline to enlarge the search somewhat, depending on the pulsar, to a range between $\delta \sim 10^{-4} - 10^{-3}$. The frequency and spin-down ranges explored in this analysis are listed in Tab. \ref{tab:tab2}.

The narrow-band search is performed using a pipeline based on the \textit{5-vector method} \cite{rob:method} and, in particular, its latest implementation, fully described in \cite{mastrogiovanni:method}, to which the reader is referred for more details. The basic idea is that of exploring a range of frequency and spin-down values by properly applying barycentric and spin-down corrections to the data in such a way that a signal would become monochromatic apart from the sidereal modulation. While a single barycentric correction applied in the time domain holds for all the explored frequency band, several spin-down corrections, one for each point in the spin-down grid, are needed. A detection statistic (DS) is then computed for each point of the explored parameter space. By using the FFT algorithm for each given spin-down value it is possible to compute the statistic simultaneously over the whole range of frequencies, this process is done for each detector, and then data is combined. The frequency/spin-down plane is then divided into frequency sub-bands ($10^{-4}$ Hz) and, for each of them, the local maximum, over the spin-down grid, of the DS is selected as a \textit{candidate}. The initial \textit{outliers} are identified among the candidates using a threshold nominally corresponding to 1\% (taking into account the number of trials\cite{rob:method}) on the p-value of the DS's noise-only distribution\footnote{The noise-only distribution is computed from the values of the DS excluded in each frequency sub-band when selecting the local maxima and then an extrapolation of the long tail of the done} and are subject to a follow-up stage in order to understand their nature. The follow-up procedure consists of the following steps: check if the outlier is close to known instrumental noise lines; compute the signal amplitude and check if it is constant throughout the run, compute the time evolution of the SNR  (which we expect to increase as the square root of the observation time for stationary noise) and compute the \textit{5-vector coherence}, which is an indicator measuring the degree of consistency between the data and the estimated waveform \cite{pia:articolo}. For each target, if no outlier is confirmed by the follow-up we set an upper-limit on the GW amplitude and NS ellipticity, see Appendix A  for more details. 
\label{sec:J1813}

\section{Selected targets}
\label{sec:targets}
We have selected pulsars whose spin-down limit could possibly be beaten, or at least approached, based on the average sensitivity of O1 data, see Fig.\ref{fig:spin_down_compar}. Pulsar distances and spin-down limits are listed in Tab. \ref{tab:tab1}. As distance estimations for the pulsars we have used the best fit value and relative uncertainties given by each indipendet measure, see pulsars list below  and Tab. \ref{tab:tab1} for more details. The uncertainty on the spin-down limit in Tab. \ref{tab:tab1} can be computed using the relation for the variance propagation\footnote{If variable $Y$ is defined from $x_i$ random variables with variance $\sigma^2_{x_i}$, then the variance $\sigma^2_{Y}$ can be estimated as: $$\sigma^2_{Y}=\sum_i \bigg( \dfrac{\partial Y}{\partial x_i}\bigg)^2 \sigma^2_{x_i} $$}.For two of these pulsars (Crab and Vela) the spin-down limit has been already beaten in a past narrow-band search using Virgo VSR4 data \cite{rob:obs}. The other targets are analysed in a narrow-band search for the first time. The timing measures for the 11 pulsars were provided by the 76-meters Lovell telescope and the 42-foot radio telescopes at Jodrell Bank (UK), the 26-meters telescope telescope at Hartebeesthoek (South Africa), the 64-meters Parkes radio telescope (Australia) and the Fermi Large Area Telescope (LAT) which is a space satellite. 
For 7 of these pulsars (Crab, Vela, J0205+6449, J1813-1246, J1952+3252, J2043 +2740 and J2229+6114) updated ephemerides covering O1 period were available and a targeted search was done in a recent work \cite{matt:O1_known} beating the spin-down limit for all of them, while for the remaining 4 pulsars we have used older measures extrapolating the rotational parameters to the O1 epoch. A list of the analysed pulsars follows:

\textbf{\textit{J0205+6449}}: Ephemerides obtained from Jodrell Bank. This pulsar had a glitch on November 11th 2015 which can affect the CW search \cite{greg:glitch}. For this reason we have performed the narrow-band search only on data before the glitch as done in \cite{matt:O1_known}. The distance  are set accordingly to \cite{kko}.

\textbf{\textit{J0534+2200 (Crab)}}: One of the high value targets for CW searches \cite{matt:O1_known} due to its large spin-down value. For this pulsar it was possible to beat the spin-down limit in a narrow-band search using Virgo VSR4 data \cite{rob:obs}. Ephemerides have been obtained from Jodrell Bank telescope \footnote{http://www.jb.man.ac.uk/pulsar/crab.html}.  
The nominal distance for the Crab pulsar and it's nebula is quoted in literature as $2.0 \pm 0.5$ kpc \cite{art:kaplan} we therefore assume the uncertainty correspond to $1\sigma$ confidence level.

\textbf{\textit{J0835-4510 (Vela)}}: Like the Crab pulsar, Vela is one of the traditional targets for CW searches. Although it spins at a relatively low frequency (compared to the others), it is very close to the Earth ($d\simeq 0.28~\mathrm{kpc}$), thus making it a potentially interesting source. Ephemerides obtained from the Hartebeesthoek Radio Astronomy Observatory in South Africa\footnote{http://www.hartrao.ac.za/}.  The distance and its uncertainty are taken accordingly to \cite{ddd}.

\textbf{\textit{J1400-6326}}: First discovered as an INTEGRAL source and then identified as a pulsar by Rossi X-ray Timing Explorer (RXTE). This NS is located in the galactic supernova remnant G310.6-1.6 and it is supposed to be quite young, the distance and its uncertainty correspond to $1\sigma$ confidence level \cite{renaud:J1400}.  

\textbf{\textit{J1813-1246}}: Ephemerides covering the O1 time span have been provided by the Fermi-LAT Collaboration\cite{matt:O1_known}. Only a lower upper-limit is present on the distance.

\textbf{\textit{J1813-1749}}: Located in one of the brightest and most compact TeV sources discovered in the HESS Galactic Plane Survey, HESS J1813-178. It is a young energetic pulsar that is responsible for the extended X-rays, and probably the TeV radiation as well. Timing obtained from Chandra and XMM Newton data \cite{halpern:J1813}, pulsar's distance and uncertainty are taken from \cite{m:messi} and correspond to $1\sigma$ confidence level.

\textbf{\textit{J1833-1034}}: Located in the Supernova remnant G21.5-0.9. This source has been known for a long time as one of the Crab-like remnants. The evidence for a pulsar was found by analysing Chandra data, the distance and its uncertainty is set accordingly to \cite{camilo:J1833} and correspond to $1\sigma$ confidence level.

\textbf{\textit{J1952+3252}}: Ephemerides have been obtained from Jodrell Bank \cite{matt:O1_known}. Distance and uncetainty taken from kinematic measures of \cite{a1b}.

\textbf{\textit{J2022+3842}}: It is a young energetic pulsar that was discovered in Chandra observations of the radio supernova remnant SNR G76.9+1.0. Distance and uncertainty are set accordingly to \cite{arumugasamy:J2022}.

\textbf{\textit{J2043+2740}}: Ephemerides obtained from the Fermi-LAT Collaboration\cite{matt:O1_known}. The distance is estimated using dispersion measure by \cite{man:cat} and using the model from \cite{yao}. The uncertainty on distance is set accordingly to  the model and correspond to $1\sigma$ confidence level.

\textbf{\textit{J2229+6114}}: Ephemerides obtained from Jodrell Bank\cite{matt:O1_known}. Distance and uncertainty are estimated by \cite{hh} using the model \cite{gg}.

\begin{table}[!h]
\caption{\label{tab:tab1}}
Distance and spin-down limit on the GW amplitude and ellipticity for the 11 selected pulsars. Distance and spin-down limit uncertainties  refer to 1$\sigma$ confidence level.
\newline
\begin{ruledtabular}
\begin{tabular}{lccc}
\textbf{Name} & \textbf{distance}[kpc] & $\mathbf{h_\mathrm{sd}} \cdot 10^{-25}$ & $\epsilon_\mathrm{sd} \cdot 10^{-4}$\\
\hline
\\
J0205+6449 \footnotemark[1] & $2.0\pm 0.3  $\footnotemark[2] & $6.9 \pm 1.1 $ & $14$\\
J0534+2200 (Crab) &$2.0 \pm 0.5 $  \footnotemark[3]& $14 \pm  3.5$  & $7.6 $\\
J0835-4510 (Vela)&  $0.28 \pm 0.02 $ \footnotemark[3]& $34\pm 2.4 $ & $18$\\
J1400-6326& $10\pm 3 $\footnotemark[4]& $0.90 \pm 0.27$ & $2.1 $ \\
J1813-1246 &$>2.5$\footnotemark[5]& $<1.8 $ & $<2.4$\\
J1813-1749 &$4.8\pm 0.3$\footnotemark[6]& $3.0 \pm 0.2 $ & $7.0 $\\
J1833-1034& $4.8 \pm 0.4 $\footnotemark[7]& $3.1 \pm 0.3  $ & $13$\\
J1952+3252& $3.0 \pm 0.5$\footnotemark[8]& $1.0 \pm 0.2 $ & $1.1 $\\
J2022+3842& $10 \pm 2 $\footnotemark[9]& $1.0 \pm 0.3$ & $6.0$\\
J2043+2740& $1.5 \pm 0.6 $\footnotemark[10]& $6.9 \pm 2.8$ & $23$\\
J2229+6114& $3.0 \pm 2$\footnotemark[3]& $3.4 \pm 2.2 $ & $6.2$\\
\end{tabular}
\end{ruledtabular}
\footnotetext[1]{This pulsar had a glitch on November 11th 2015}
\footnotetext[2]{Distance from neutral Hydrogen absorption of pulsar wind nebula 3C 58 \cite{kko}}
\footnotetext[3]{Distance taken from independent measures reported in ATNF catalog, see text for references}
\footnotetext[4]{Distance from dispersion measures \cite{renaud:J1400}}
\footnotetext[5]{Lower limit of \cite{m:mar}}
\footnotetext[6]{Distance from Chandra and XMM-Newton  from \cite{halpern:J1813}}	
\footnotetext[7]{Distance from Parkes telescope \cite{camilo:J1833}}
\footnotetext[8]{Distance from kinematic distance of the associated supernova remnant \cite{a1b}}
\footnotetext[9]{Distance of the hosting supernova remnant \cite{arzoumanian:J2022}. In some papers a distance value of $\sim$10 kpc is considered \cite{arumugasamy:J2022}.}
\footnotetext[10]{Distances taken from v1.56 of the ATNF Pulsar Catalog\cite{man:cat}}

\end{table}

\section{Results}
\label{sec:results}
In this section we discuss the results of the analysis. First, in Sec. \ref{sec:cand} we briefly describe the initial outliers, for most of which the follow-up described in \ref{sec:method} has been enough to exclude a GW origin. Two  outliers, belonging respectively, to the Vela and J1833-1034 pulsars needed a deeper study. The studies discussed in detail in the next section, disfavour  the signal hypothesis and seem to suggest these outliers as marginal noise events. Nevertheless the outliers showed some promising features and for this reason a follow-up using O2 data has been carried out and described in Appendix C.  The outliers were no longer present in O2 data and therefore inconsistent with persistent CW signals. Finally, in Sec. \ref{sec:upper} upper limits on the strain amplitude for the eleven targets are discussed.

\subsection{Outliers outlook}
\label{sec:cand}
\begin{figure}[H]
\includegraphics[width=0.45\textwidth]{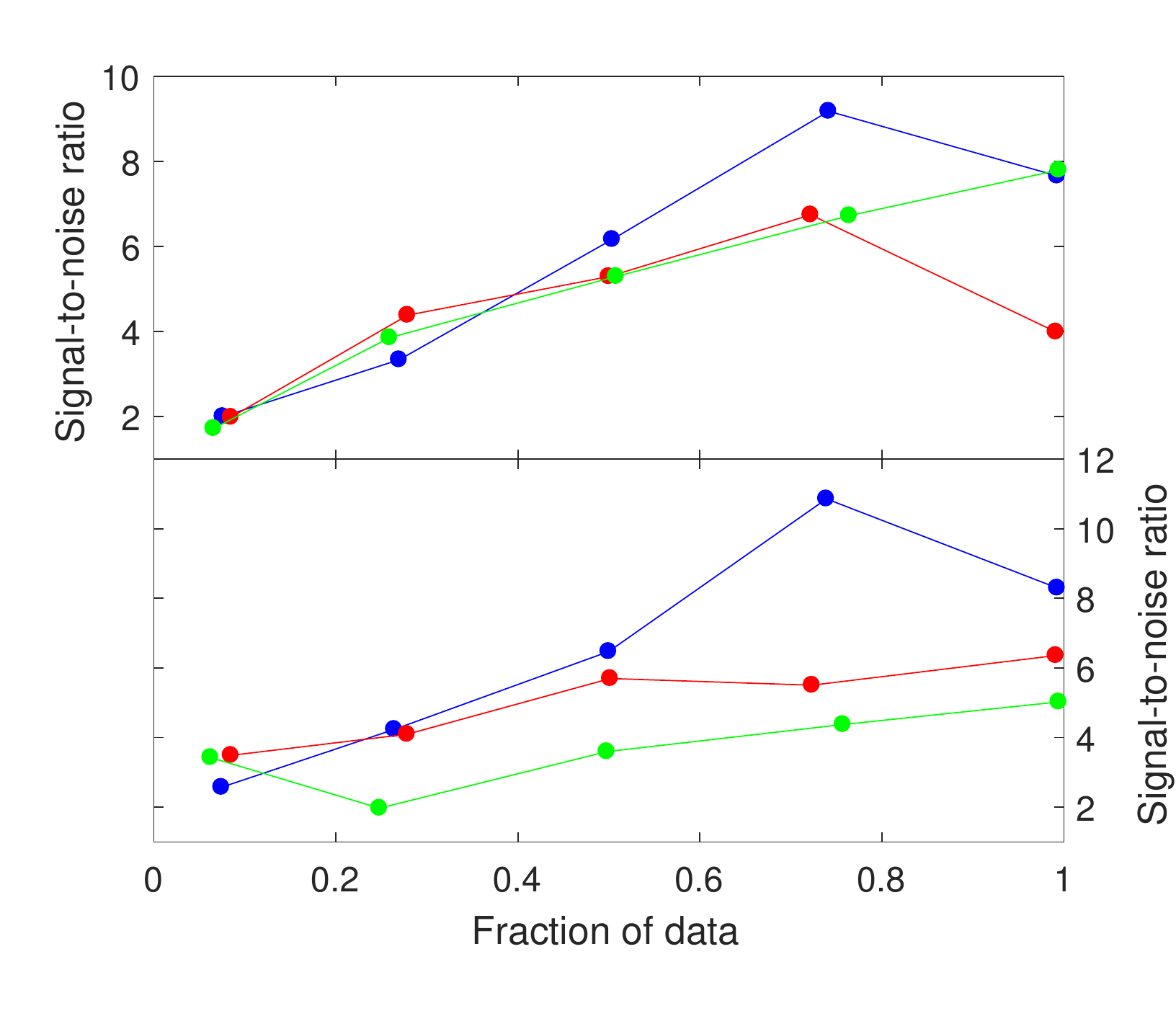}
\caption{Top panel: SNR  computed with respect to the fraction of data for J1833-1034 outlier in  Hanford (red line), Livingston (green) and joint (blue) analysis respectively. Bottom panel: SNR  computed with respect to the fraction of data for Vela outlier in  Hanford (red line), Livingston (green) and joint (blue) analysis respectively.}
\label{fig:SNR}
\end{figure}

We have found initial outliers for 9 of the 11 analysed pulsars. More precisely, for most pulsars we have found one or two outliers, with the exception of J1813-1749 (36 outliers) and J1952+3252 (6 outliers). For J2043+2740 and J2229+6114 no outlier has been found. A summary of the outliers found in the analysis is given in Tab. \ref{tab:tab_cand}. The follow-up has clearly shown that in the case of J1952+3252 and J1813-1749 the outliers arise from noise disturbances in LIGO H (for J1813-1749) and in LIGO L (for J1952+3252), see Appendix B for more details. Most of the remaining outliers show an inconsistent time evolution of the SNR  together with a low coherence between LIGO H and LIGO L and hence have been ruled out.  As mentioned before, two outliers, one for J1833-1034 and one for Vela, have shown promising features during the basic follow-up steps: no known noise line is present in their neighborhood, the amplitude estimation is compatible and nearly constant among the LIGO L and LIGO H runs and their SNR  appears to increase with respect to the integration time (see Fig. \ref{fig:SNR}). Even if the trend of the SNR does not increase monotonically with time, as expected for real signals, we have decided to follow-up this outliers due to the fact that they show a completely different SNR trend with respect to all the other outliers found in this work. Moreover each outlier's significance increases in the multi-IFOs search, suggesting a possible coherent source.

\paragraph*{J1833-1034 and Vela outliers:} In order to establish  if the outliers were not artefacts created by the narrow-band search, we  also looked for the two outliers using two other analysis pipelines for targeted searches, which used a Bayesian approach: one designed for searching for non-tensorial modes in CW signals \cite{isi:nGR}, the other developed for canonical CW target searches \footnote{frequency and spin-down value fixed to the outlier's value found in the narrow-band search}  and parameter estimation \cite{matt:bayes}. Both pipelines produced odds, listed in Tab \ref{tab:odd}, which show a small preference for the presence of a candidate compatible with general relativity. The odd values are not surprising due to the fact that we are using values for the frequency and the spin-down which are fixed to the ones found in the narrow-band search. Hence, a trial factor should be taken into account in order to make a robust estimation on the signal hypothesis preference. Besides the previous considerations, the values in Tab. \ref{tab:odd} clearly shows that the outliers are not artefacts created by the narrow-band pipeline. We have also compared the estimation of the outlier parameters obtained from the {\it 5-vector}, $\mathcal{F}$-statistic and Bayesian\cite{pia:articolo, krolak:fstat, matt:bayes} pipelines.  The inferred parameters are listed in Tab.\ref{tab:out_par} and seems to be compatible among the three independently developed targeted pipelines, thus suggesting the true presence of these outliers inside the data.

\begin{table}[H]
\caption{\label{tab:odd}}
Odds obtained for the two outliers by the Bayesian pipelines \cite{isi:nGR, isi:LVC}. The second column shows the odds of any non-tensorial signal hypothesis versus the canonical CW signal hypothesis, the third column is the odds ratio of the canonical signal hypothesis vs the gaussian noise hypothesis while the last column is the odds ratio between the coherent signal among the two detectors vs  the hypothesis that the outliers arise from an incoherent noise between LIGO H and L.   
\begin{ruledtabular}
\begin{tabular}{lccc}
\textbf{Name} & $\log_{10} \mathcal{O}^\mathrm{\text{nGR}}_\mathrm{GR}$ & $\log_{10} \mathcal{O}^\mathrm{S}_\mathrm{N}$ & $\log_{10} \mathcal{O}^{\text{C}}_\mathrm{I}$ \\
\hline
J0835-4510 (Vela)& $-0.55$& $2.30$ & $1.07$ \\
J1833-1034 & $-0.73$& $2.73$ & $1.34$\\
\end{tabular}
\end{ruledtabular}
\end{table}
\begin{table}[H]
\caption{\label{tab:out_par}}
Estimation of the GW parameters, $h_0$, $\cos \iota$ and $\psi$, from three targeted search pipelines  \cite{pia:articolo, krolak:fstat, matt:bayes}. The intervals refer to the 95\% confidence level.
\begin{ruledtabular}
\begin{tabular}{lcccc}
\textbf{J0835-4510 (Vela)} & $h_0 \cdot 10^{-25}$ & $\cos \iota$ & $\psi$[rad] \\
\hline \\
5-vector & $5.7^{+2.3}_{-2.1}$&\centering$-0.09^{+0.27}_{-0.19}$ & $0.69^{+0.57}_{-0.58}$ \\
Bayesian & \centering$6.6^{+3.1}_{-3.7}$&\centering $-0.14^{+0.28}_{-0.48}$ & $0.57^{+0.31}_{-0.30}$ \\
$\mathcal{F}$-statistic  & \centering$7.1$&\centering $-0.13$ & $0.55$ \\
\hline \hline
\textbf{J1833-1034} & $h_0 \cdot 10^{-25}$ & $\cos \iota$ & $\psi$[rad] \\
\hline \\
5-vector &\centering $1.6^{+0.5}_{-0.6}$&\centering $0.10^{+0.30}_{-0.20}$ & $0.58^{+0.35}_{-0.51}$\\
Bayesian & \centering$1.8^{+0.8}_{-1.7} $& \centering$0.24^{+0.64}_{-0.31}$ & $0.58^{+0.56}_{-0.51}$\\
$\mathcal{F}$-statistic & \centering$2.0$& \centering$0.22$ & $0.59$
\\
 
\end{tabular}
\end{ruledtabular}
\end{table}

\begin{figure}[H]
\includegraphics[width=0.5\textwidth]{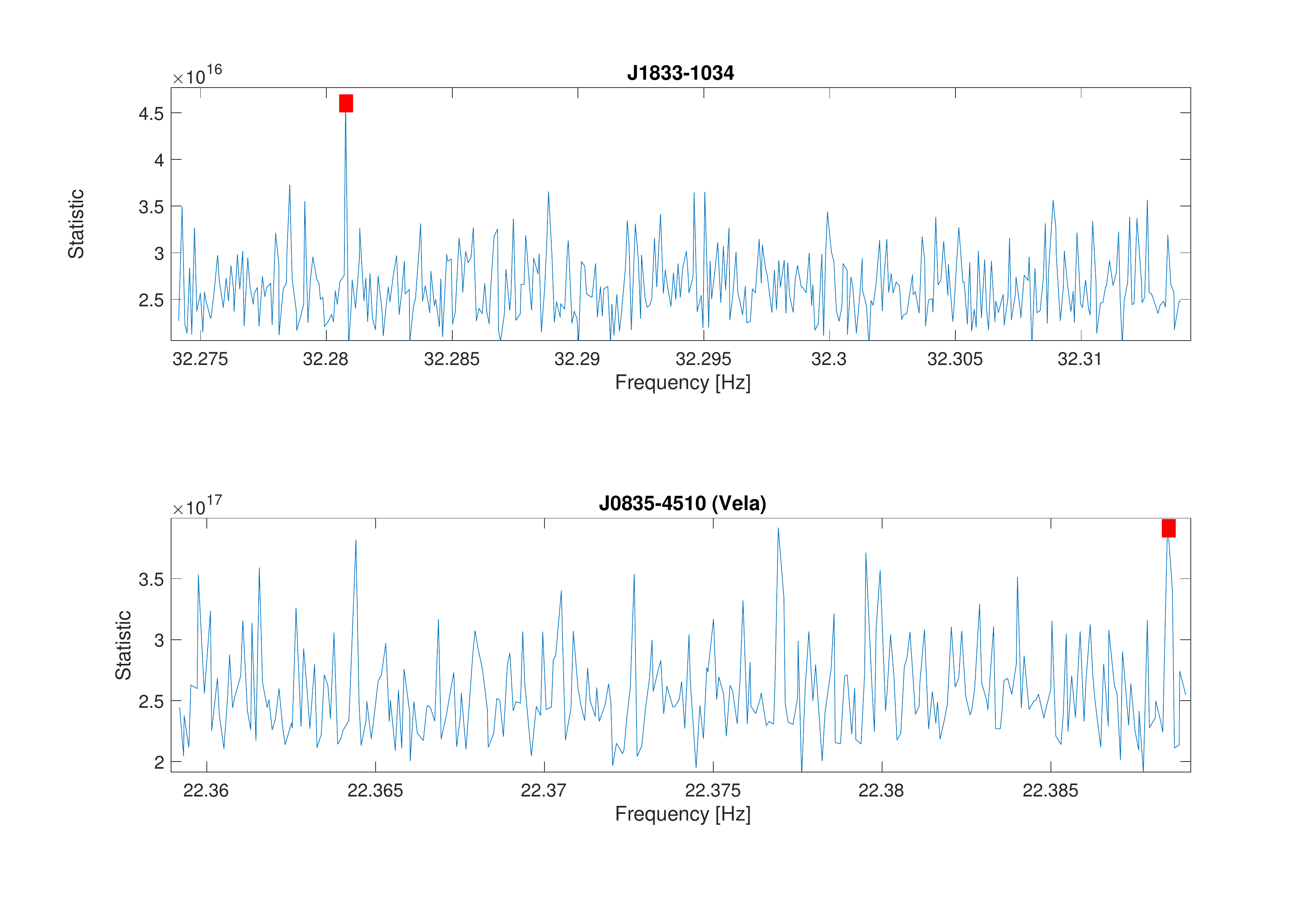}
\caption{Values of the local maximum of the DS over the spin-down corrections and the frequency sub-bands for J1833-1034 (top panel) and Vela (bottom panel). The outliers are highlighted with the red square.}
\label{fig:DS_18}
\end{figure}

 In order to establish each outlier's nature, a complete understanding of the noise background  is needed. For this reason the first check was to look at the DS distribution in the narrow-band search. In the presence of a true signal we expect to see a single significant peak in the DS. Figure \ref{fig:DS_18} shows the distribution of the DS (maximized over the spin-down corrections) for J1833-1034 and for Vela over the frequency band analysed. We notice that for J1833-1034 the outlier is the only clear peak present in the analysis, surrounded by several lower peaks in the detection statistic which are not above the corresponding p-value threshold. On the other hand, for Vela, several peaks in the DS are present, with significance below but similar to that of the outlier, thus suggesting that the Vela outlier can be due to non-gaussian background. 
 
\begin{figure}[H]
\includegraphics[width=0.5\textwidth]{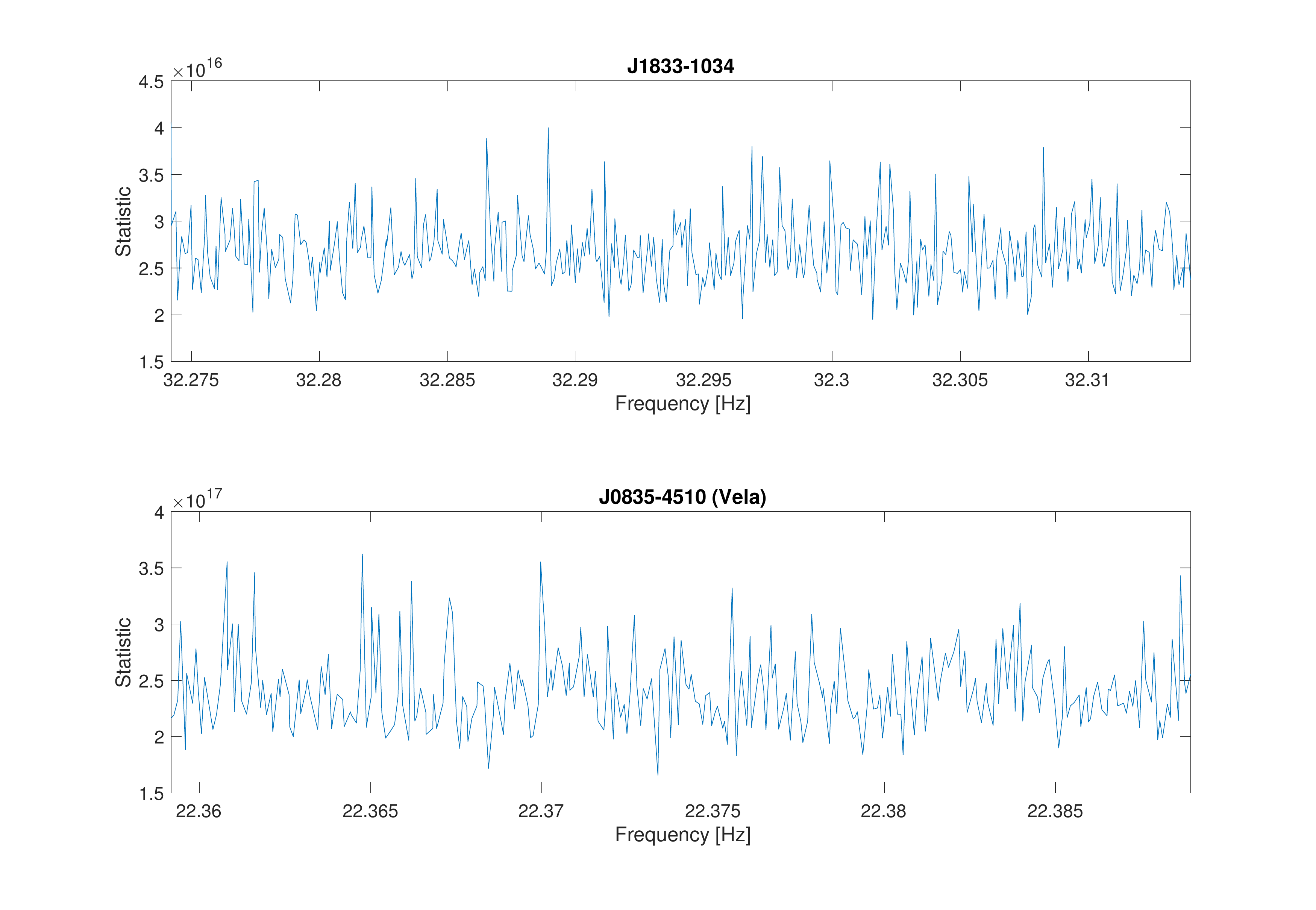}
\caption{Values of the local maximum of the DS over the spin-down corrections and the frequency sub-bands for a close sky position to J1833-1034 (top panel) and Vela (bottom panel).}
\label{fig:nb_ds}
\end{figure}

 A further test consist of checking the distribution of the DS in a narrow-band search performed using the same frequency/spin-down region but in a sky-position shifted by about 0.5 degrees. Using this method we keep the contribution of non-Gaussian noise in the DS while removing a possible signal contribution. Figure \ref{fig:nb_ds}  shows the distribution of the DS obtained for J1833-1034 and Vela outliers.  In both cases no over-threshold peak are present, however the analysed bands seem similarly polluted by non-Gaussian contributions which produce peaks in the DS. 
We have also studied the significance of the outliers using two of the three targeted search pipelines. As done previously, we have built a noise-distribution of the DS, performing the targeted searches in other sky positions in order to compute the outliers p-value. Using the trials factor from the narrow-band search we have found the outliers to have a higher resulting p-value with respect to the 1\% threshold used in the initial outliers selection process during the narrow-band search,  increasing the likelihood that these outliers were generated from noise.  
Some of the  previous tests disfavour the signal hypothesis and seem to indicate the presence of a coherent noise disturbance among the interferometers. Previous works such as \cite{matt:O1_known,eath:all}  have already pointed out the presence of some non-trivial coherent noise artefacts among the IFOs which can produce outliers. For this reason, in the spirit of what is done in \cite{eath:all}, we have looked at O2 data.  If the outliers are really due to a ``standard'' CW signal, they are expected to be present also in O2 data, due to their persistent nature. We have analysed the data using the narrow-band pipeline but no evidence for these outliers was found in data. In conclusion the outliers are not true CW signals. More details on the O2 analysis can be found in Appendix C.

\subsection{Upper limits}

\label{sec:upper}
Following the procedure described in Appendix A we have set  95\%  C.L. upper-limits on GW strain amplitude in every $10^{-4} \text{Hz}$ sub-band. In each of these bands the upper-limit was computed by injecting simulated GW signals with several different amplitudes and finding the amplitude such that 95\% of the injected signals with that amplitude produce a value of the DS  corresponding to the nominal overall p-value of 1\%.  Tab. \ref{tab:tab3} gives an overview of the overall sensitivity reached in our search using the median of the upper-limits among the analysed frequency band: graphs of the upper-limits see Fig. \ref{fig:datahist}. For J2043+2740, J1952+3252 and J2022+3842 our overall sensitivity is clearly above the spin-down limit. For J1813-1246 and J1833-1034 our overall sensitivity is close to the spin-down limit, producing values of the upper-limits both below and above the spin-down limit. For J1400-6326 we have obtained a large fraction of the upper-limits in the narrow-band search below the spin-down while for  J0205+6449 and  J2229+6114 we have beaten the spin-down limit in a narrow-band search for the very first time.
For Crab and Vela pulsars we have obtained upper-limits respectively $\sim$ 7 and $\sim$3.5 times lower than those computed in a past analysis \cite{rob:obs}. This improvement is due to a combination of two factors: the enhanced sensitivity of advanced detectors and the choice to compute upper limits over $10^{-4}$ Hz sub-bands instead of the full analysis band, thus reducing the impact of the look-elsewhere effect in each sub-band \cite{rob:method}. Finally the narrow-band search for J1813-1749 beats the spin-down limit (if we exclude from the search the frequency region around the LIGO H artefact), constraining for the first time their CW emission. Pulsars J1813-1749 and J1400-6326 have not been previously analysed in targeted searches, due to the lack of ephemeris covering O1 or previous runs. 
Even if we consider the uncertainties on the pulsars distance, propagated in Tab. \ref{tab:tab3} for the spin-down limit and upper-limit ratio, we are still able to beat the spin-down for those 5 pulsars. 
\begin{table}[H]
\caption{\label{tab:tab3} Median over the analysed frequency band of the upper-limits obtained on the GW amplitude for the 11 known pulsars. In the fourth column we report the ratio between the spin-down limit listed in Tab. \ref{tab:tab1} and the median of the upper-limit, uncertainties correspond to $1\sigma$ confidence level and are due to the uncertainties on pulsars distance. The last column reports the median upper-limit on the fraction of rotational energy lost due to GW emission.}\begin{ruledtabular}
\begin{tabular}{lcccc}
\textbf{Name} &$h_{\text{ul}}  $& $\epsilon_{\text{ul}} $ & $h_{\text{ul}} / h_\mathrm{sd}$ & $\dot{E}_\mathrm{rot} / \dot{E}_\mathrm{GW}$\\
&$\cdot 10^{-25}$& $\cdot 10^{-4}$ &  & \\
\hline
J0205+6449& $3.76 $& $7.7 $& $0.54 \pm 0.09$ &$0.29$\\
J0534+2200 (Crab) &$1.08 $& $0.58$ & $0.07 \pm 0.02$ &$0.005$ \\
J0835-4510 (Vela) & $9.28 $& $5.3 $& $0.27 \pm 0.02 $&$0.07$\\
J1400-6326& $1.17$& $2.7 $& $1.3 \pm 0.4 $&-\\
J1813-1246 &$1.80 $& $2.5 $& $>1.0$&-\\
J1813-1749 &$1.9 $& $4.8 $& $0.64 \pm 0.04 $&$0.41$\\
J1833-1034& $3.08 $& $13 $& $0.99 \pm 0.09 $&-\\
J1952+3252& $1.31$& $1.4 $& $1.31 \pm 0.22 $&-\\
J2022+3842& $1.90 $& $11 $& $1.77 \pm 0.35 $&-\\
J2043+2740& $14.4 $& $47 $& $2.07 \pm 0.83 $&-\\
J2229+6114& $1.78 $& $3.4$& $0.54 \pm 0.35$&$0.30$\\
\end{tabular}
\end{ruledtabular}
\end{table}
\section{Conclusion}
\label{sec:conclusion}


In this paper we have reported the result of the first narrow-band search using Advanced LIGO O1 data for 11 known pulsars. For each pulsar, a total of about $10^7$ points in the frequency and spin-down space have been explored. For 9 pulsars, outliers have been found and analysed in a follow-up stage. Most of the outliers did not pass the follow-up step and were labeled as noise fluctuations or instrumental noise artefacts. We have found two near-threshold outliers, one for J1833-1034 and another for the Vela pulsar, which needed deeper studies but eventually  were rejected. In particular, the  outliers have been searched for in the first five months of LIGO O2 run and were not confirmed. We have computed upper-limits on the signal strain, finding for 5 pulsars values below the spin-down limit in the entire narrow-band search (Crab, J1813-1749, J0205+6449, 2229+6114 and Vela). For the Crab and Vela pulsars the upper limits significantly improve with respect to past analyses. For an additional 3 targets (J1833-1034, J1813-1246 and J1400-6326), the median upper limit across the search bands is below or very close the spin-down limit. For J1813-1749, which have never been analysed in a targeted search, we have beaten the spin-down limit for the first time while for  J0205+6449 and J2229+6114 the spin-down limit has been beaten for the first time in a narrow-band search. 
7 of the 11 pulsars analysed in this work, were also analysed using O1 data in a target search \cite{matt:O1_known}. The upper-limits found in this work are about 2-3 times higher with respect to targeted searches: the sensitivity loss is due to the fact that we are exploring a large number of templates in the frequency spin-down plane.  On the other hand we have put for the first time upper-limits in a small frequency spin-down region around the expected values.

The analysis of forthcoming Advanced LIGO and Virgo runs \cite{lvc:pros}, with improved sensitivities and longer durations, could provide the first detection of continuous gravitational signals  from spinning neutron stars, which would help to shed light on their structure and properties.
\section*{Acknowledgments}

The authors gratefully acknowledge the support of the United States
National Science Foundation (NSF) for the construction and operation of the
LIGO Laboratory and Advanced LIGO as well as the Science and Technology Facilities Council (STFC) of the
United Kingdom, the Max-Planck-Society (MPS), and the State of
Niedersachsen/Germany for support of the construction of Advanced LIGO 
and construction and operation of the GEO600 detector. 
Additional support for Advanced LIGO was provided by the Australian Research Council.
The authors gratefully acknowledge the Italian Istituto Nazionale di Fisica Nucleare (INFN),  
the French Centre National de la Recherche Scientifique (CNRS) and
the Foundation for Fundamental Research on Matter supported by the Netherlands Organisation for Scientific Research, 
for the construction and operation of the Virgo detector
and the creation and support  of the EGO consortium. 
The authors also gratefully acknowledge research support from these agencies as well as by 
the Council of Scientific and Industrial Research of India, 
Department of Science and Technology, India,
Science \& Engineering Research Board (SERB), India,
Ministry of Human Resource Development, India,
the Spanish  Agencia Estatal de Investigaci\'on,
the  Vicepresid\`encia i Conselleria d'Innovaci\'o, Recerca i Turisme and the Conselleria d'Educaci\'o i Universitat del Govern de les Illes Balears,
the Conselleria d'Educaci\'o, Investigaci\'o, Cultura i Esport de la Generalitat Valenciana,
the National Science Centre of Poland,
the Swiss National Science Foundation (SNSF),
the Russian Foundation for Basic Research, 
the Russian Science Foundation,
the European Commission,
the European Regional Development Funds (ERDF),
the Royal Society, 
the Scottish Funding Council, 
the Scottish Universities Physics Alliance, 
the Hungarian Scientific Research Fund (OTKA),
the Lyon Institute of Origins (LIO),
the National Research Foundation of Korea,
Industry Canada and the Province of Ontario through the Ministry of Economic Development and Innovation, 
the Natural Science and Engineering Research Council Canada,
Canadian Institute for Advanced Research,
the Brazilian Ministry of Science, Technology, Innovations, and Communications,
International Center for Theoretical Physics South American Institute for Fundamental Research (ICTP-SAIFR), 
Russian Foundation for Basic Research,
Research Grants Council of Hong Kong,
the Leverhulme Trust, 
the Research Corporation, 
Ministry of Science and Technology (MOST), Taiwan
and
the Kavli Foundation.
The authors gratefully acknowledge the support of the NSF, STFC, MPS, INFN, CNRS and the
State of Niedersachsen/Germany for provision of computational resources.
 Pulsar observations with the Lovell telescope and their analyses are supported through a consolidated grant from the STFC in the UK. The Nan\c{c}ay Radio Observatory is operated by the Paris Observatory, associated with the French CNRS.

\setcounter{table}{5}

\twocolumngrid

\onecolumngrid
\setcounter{table}{4}

\begin{table}[htpb!]
\caption{\label{tab:tab2} This table reports the explored range for the rotational parameters of each pulsar. The columns are: the central frequency of the search ($f_0$), explored frequency band ($\Delta f$), central spin-down value of the search ($\dot{f}_0$), explored spin-down band ($\Delta \dot{f}_0$), the number of frequency bin explored ($n_f$), and the number of spin-down values explored ($n_{\dot{f}}$). All the rotational parameters are scaled at the common reference time on September 12th 2015.}
\begin{ruledtabular}
\begin{tabular}{lccccccc}
\textbf{Name}&$\mathbf{f_0}$ [Hz]&$\mathbf{\Delta f}$ [Hz]&$\mathbf{\dot{f}_0}$ [Hz/s]&$\mathbf{\Delta \dot{f}}$ [Hz/s] & $\mathbf{n_f}$  &$\mathbf{n_{\dot{f}}}$ \\ \hline
J0205+6449 &$30.4095820$& $0.03$& $-8.9586\cdot 10^{-11}$& $1.75\cdot 10^{-13}$& $2.5 \cdot10^6$& $19$\\
J0534+2200 (Crab) &$59.32365204$& $0.10$& $-7.3883 \cdot 10^{-10}$& $1.48 \cdot 10^{-12}$& $18.5\cdot10^6$& $161$\\
J0835-4510 (Vela)& $22.3740981$& $0.03$& $-3.1191\cdot 10^{-11}$& $6.43\cdot 10^{-14}$& $2.5 \cdot10^6$& $7$\\
J1400-6326& $64.1253722$& $0.07$& $-8.0017\cdot 10^{-11}$& $1.75\cdot 10^{-13}$& $6.5 \cdot10^6$& $19$\\
J1813-1246 &$41.6010333$& $0.04$& $-1.2866\cdot 10^{-11}$& $6.43\cdot 10^{-14}$& $3.4 \cdot10^6$& $7$\\
J1813-1749 &$44.7128464$& $0.05$&$ -1.5000\cdot 10^{-10}$& $3.03\cdot 10^{-13}$& $2.5 \cdot10^6$& $33$\\
J1833-1034& $32.2940958$& $0.04$&$ -1.0543\cdot 10^{-10}$& $2.11\cdot 10^{-13}$& $3.4 \cdot10^6$& $23$\\
J1952+3252& $50.5882336$& $0.05$& $-7.4797\cdot 10^{-12}$& $6.43\cdot 10^{-14}$& $4.3 \cdot10^6$& $7$\\
J2022+3842& $41.1600845$& $0.04$& $-7.2969\cdot 10^{-11}$ & $1.60\cdot 10^{-13}$& $3.4 \cdot10^6$& $17$\\
J2043+2740& $20.8048628$& $0.05$& $-3.4390\cdot 10^{-11}$& $6.43\cdot 10^{-14}$& $4.3 \cdot10^6$& $7$\\
J2229+6114& $38.7153156$& $0.06$& $-5.8681\cdot 10^{-11}$& $1.19\cdot 10^{-13}$& $5.1 \cdot10^6$& $13$\\
\end{tabular}
\end{ruledtabular}
\end{table}
\setcounter{table}{5}

\begin{table}[H]
\caption{\label{tab:tab_cand} The table reports the outliers found in our analysis for each analysed pulsar. The first column is the name of pulsar, the second indicates the number of outliers found in the analysis. The third and the fourth columns show respectively the outier frequency and spin-down. The last column reports the corresponding p-value. For the two targets J1813-1749 and J1952+3252 the outliers did not undergo the follow-up procedure due to the fact that can easily associated with known noise lines, see Appendix B.}
\begin{ruledtabular}
\begin{tabular}{lcccc}
\textbf{Name}&N. of candidates&Frequency [Hz]&Spin-down[Hz/s]&P-value \\ \hline
J0205+6449 &1&$30.4046480$& $-8.937 \cdot 10^{-11}$& $0.003$\\
J0534+2200 (Crab) &1&$59.3702101$& $-7.3920 \cdot 10^{-10}$& $0.005$\\
J0835-4510 (Vela) &1&$22.3884563$& $-3.12 \cdot 10^{-12}$& $0.009$\\
J1813-1246 &2&$41.5779102,41.5852264$& $-1.285 \cdot 10^{-11},-1.284 \cdot 10^{-11}$& $0.007,0.005$\\
J1813-1749& 36 &close to $44.705$ Hz&-&$<10^{-6}$\\
J1833-1034 &1&$32.2807633$& $-1.0535 \cdot 10^{-10}$& $0.0004$\\
J1952+3252&6 &close to $50.601$&-&$<10^{-5}$\\
J1400-6326&2 &$64.1089253,64.1406011$& $-8.008 \cdot 10^{-11},-8.937 \cdot 10^{-11}$& $0.002,0.003$\\
J2022+3842 &1&$41.1603319$& $-7.297 \cdot 10^{-11}$& $0.007$\\
\end{tabular}
\end{ruledtabular}
\end{table}

\setcounter{figure}{5}

\begin{figure}[H]
\includegraphics[width=0.329\textwidth]{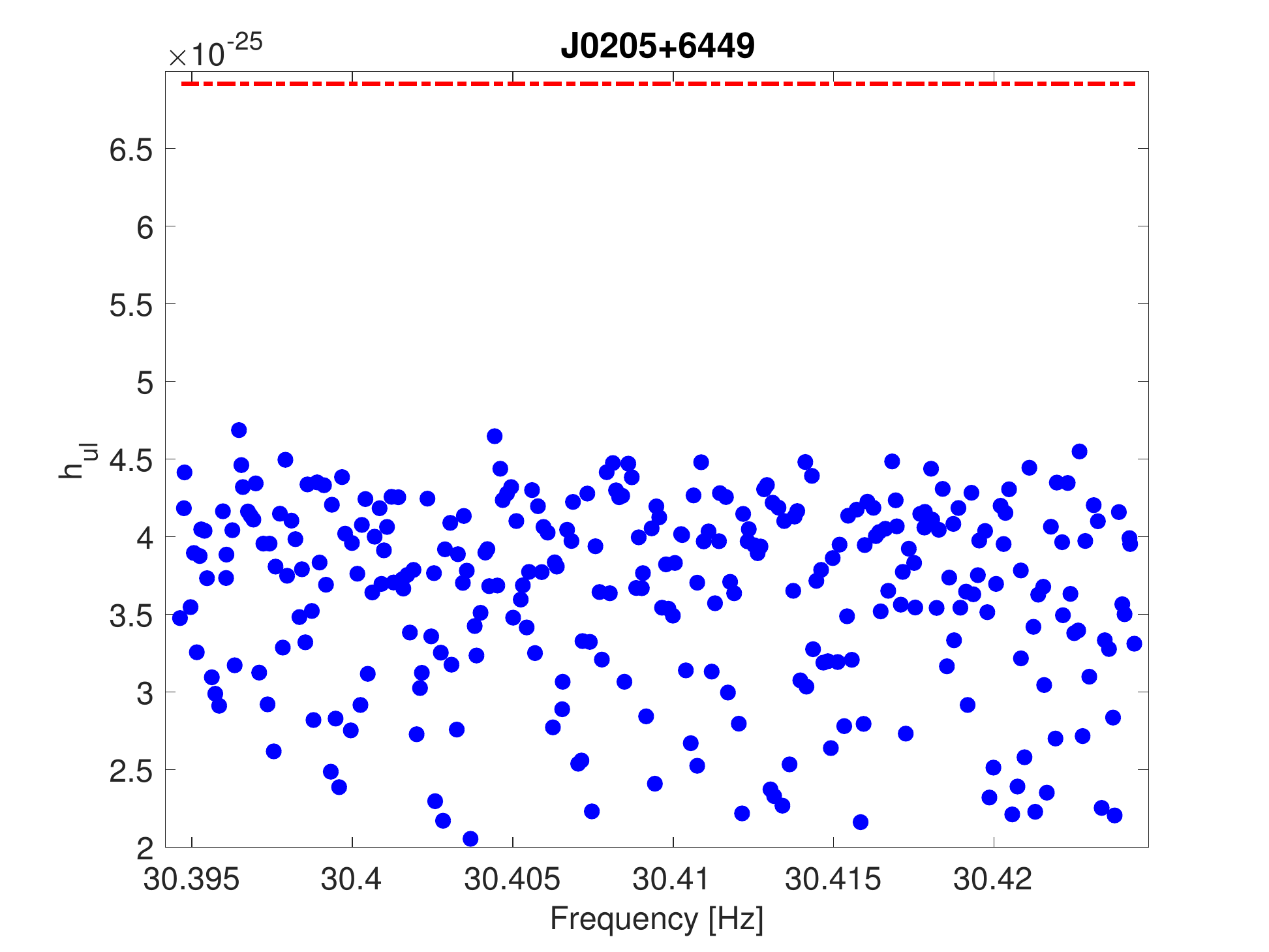}
\includegraphics[width=0.329\textwidth]{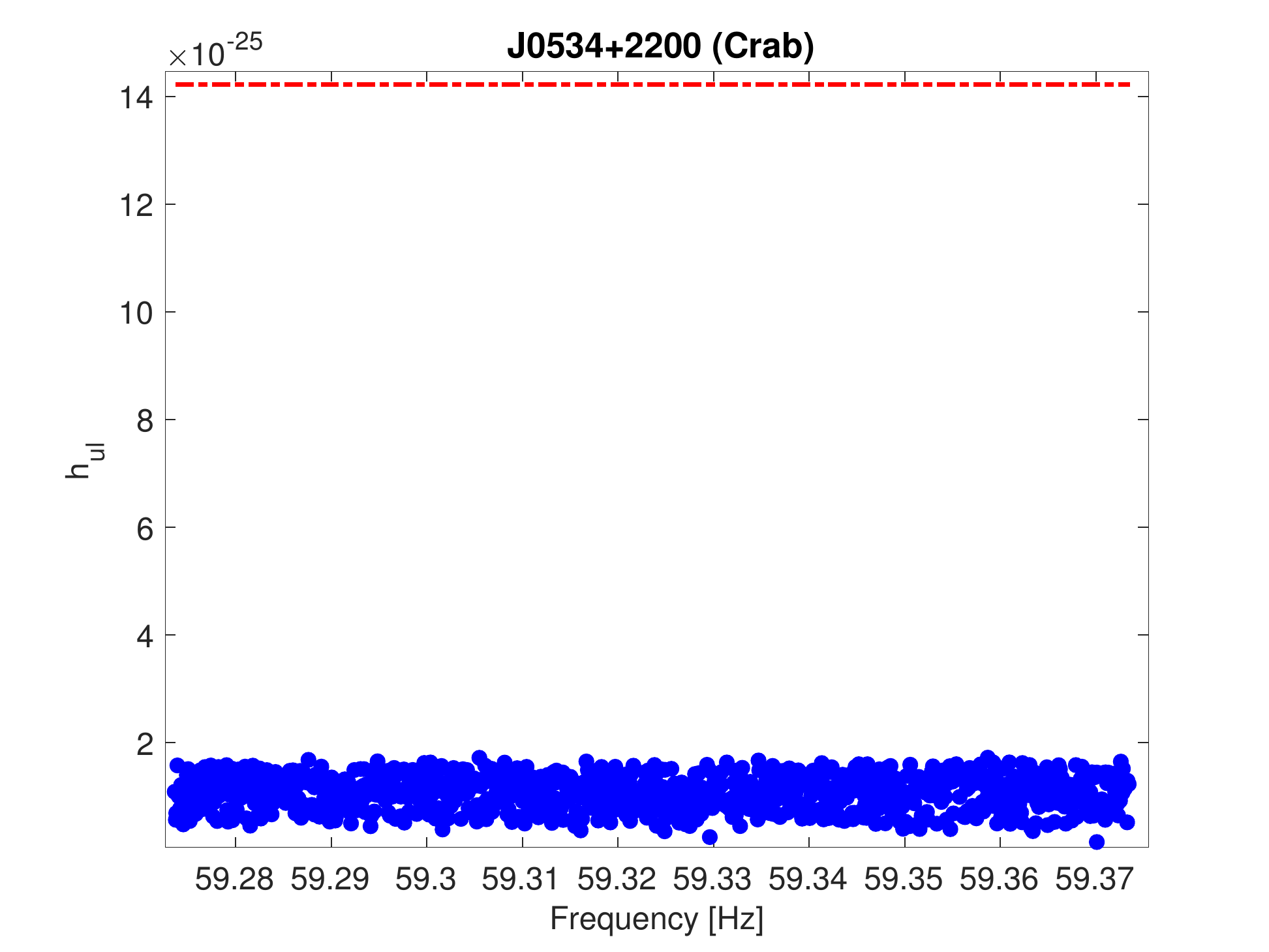}
\includegraphics[width=0.329\textwidth]{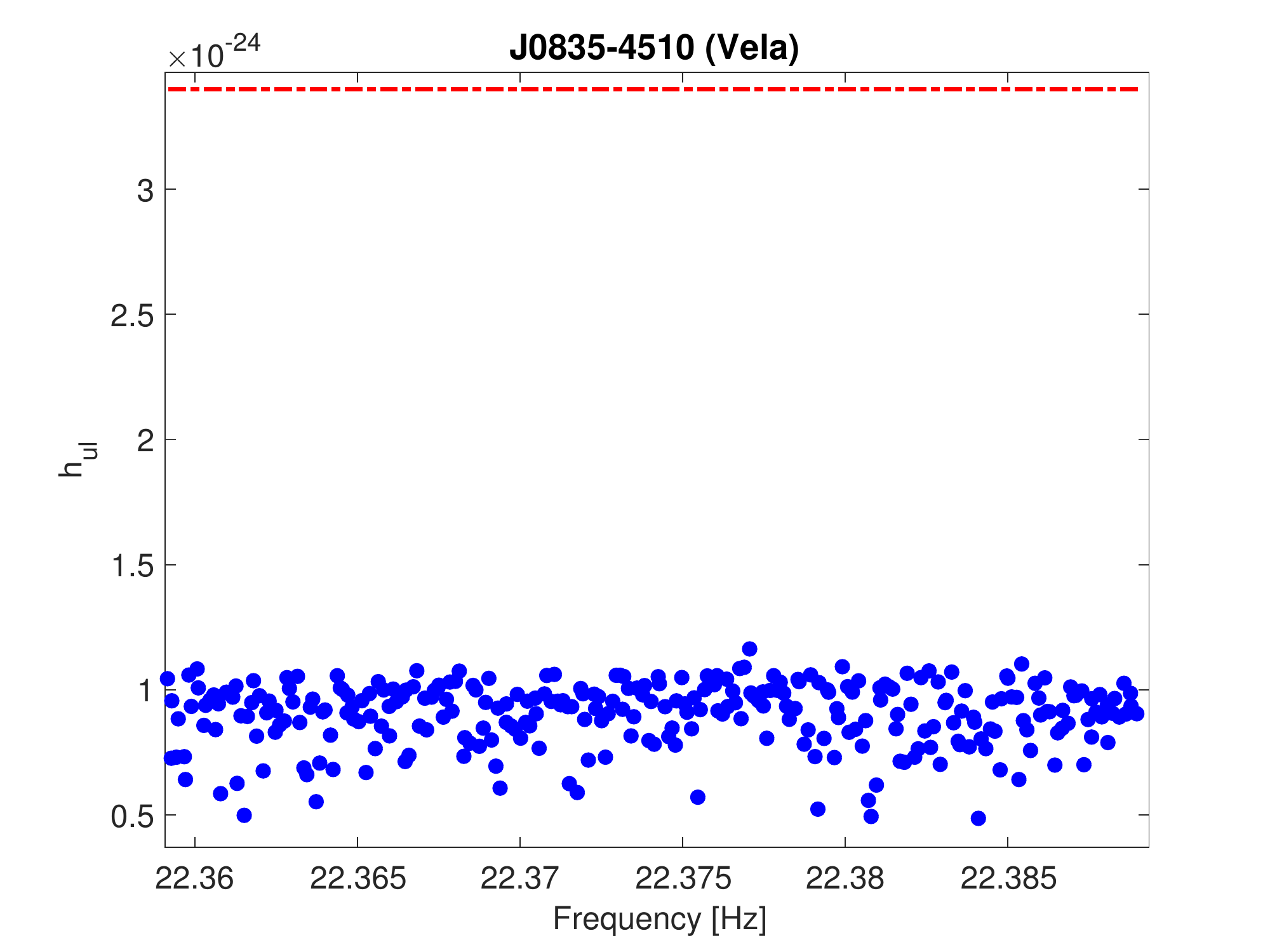}
\includegraphics[width=0.329\textwidth]{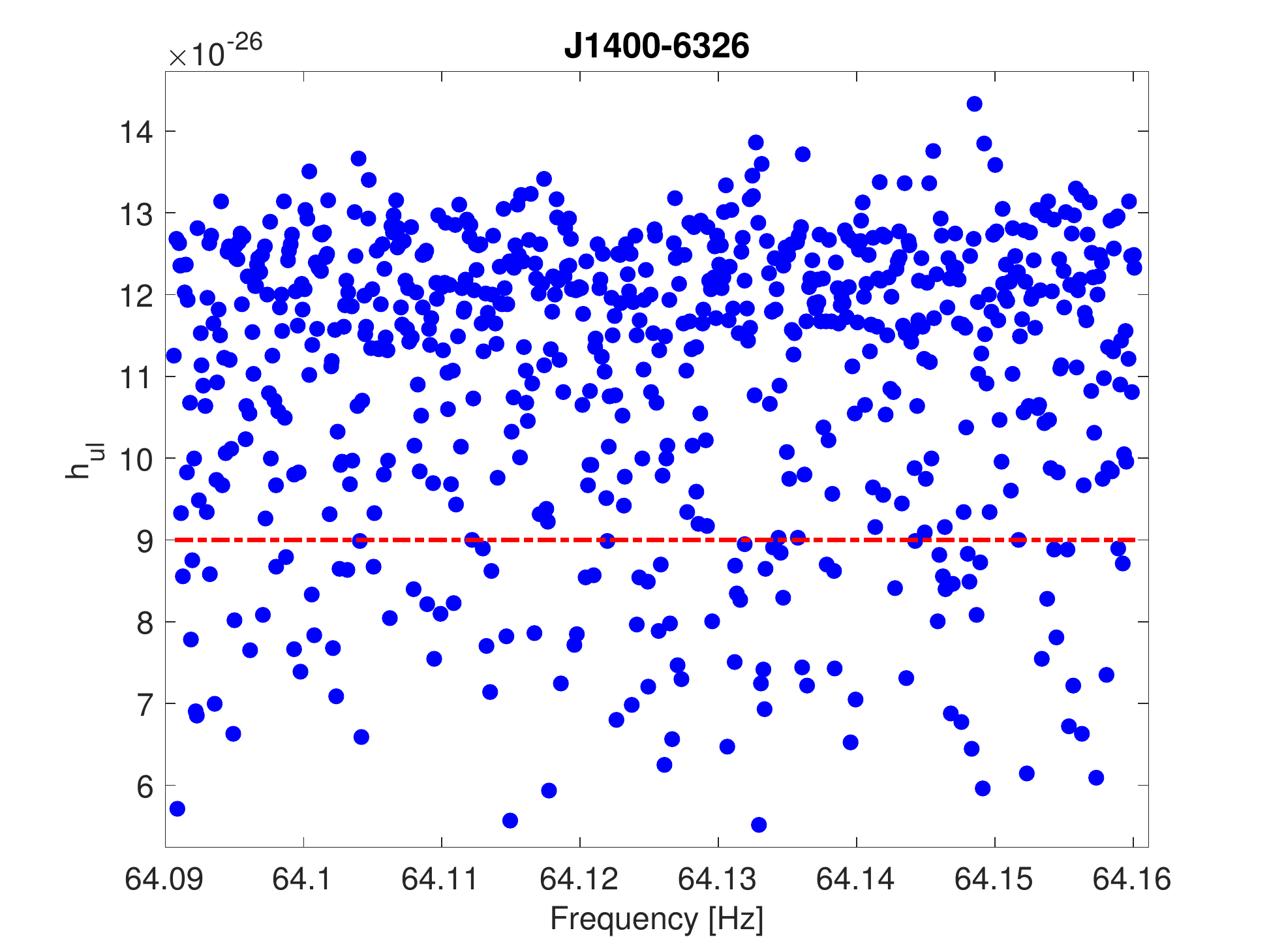}
\includegraphics[width=0.329\textwidth]{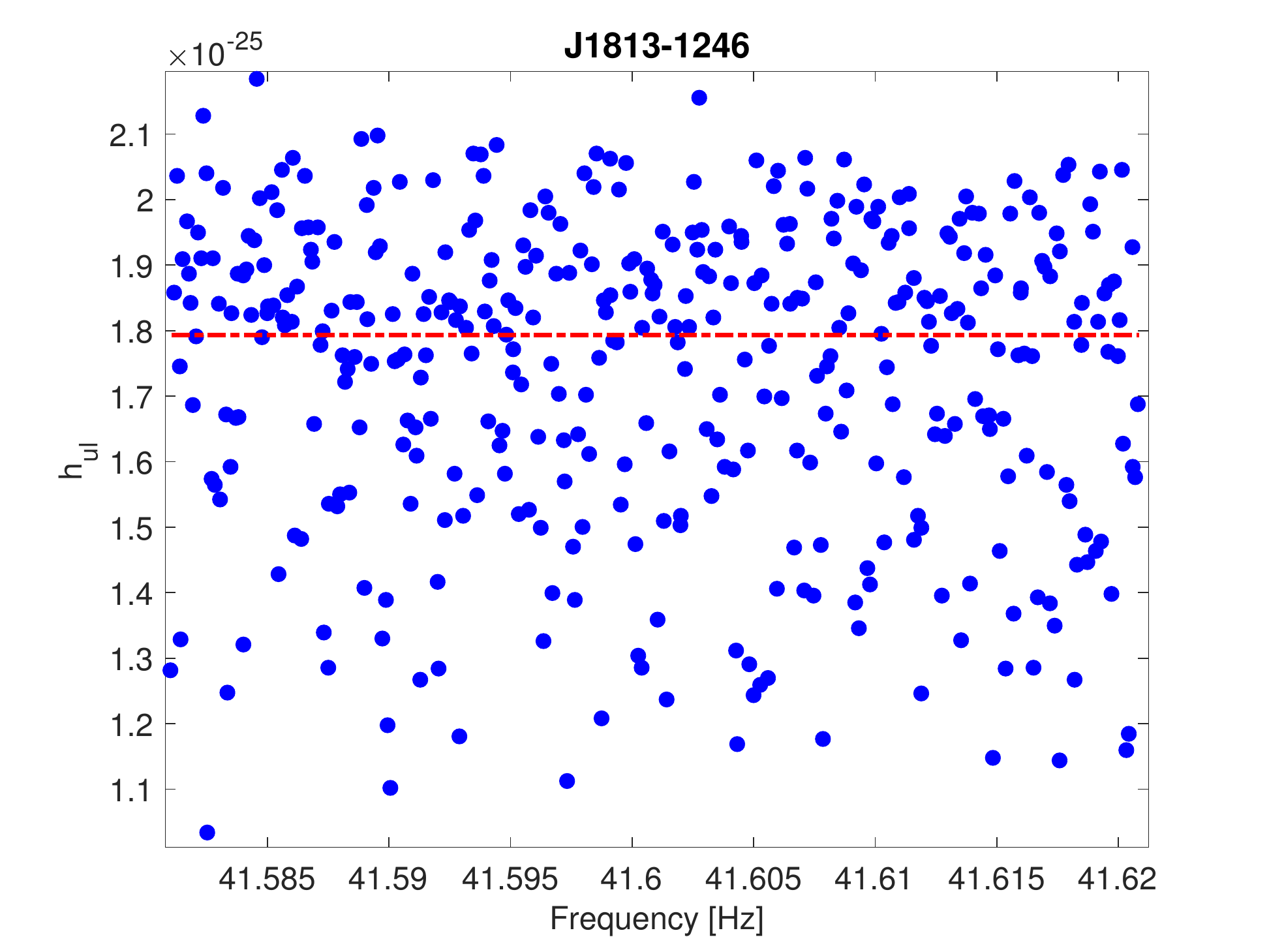}
\includegraphics[width=0.329\textwidth]{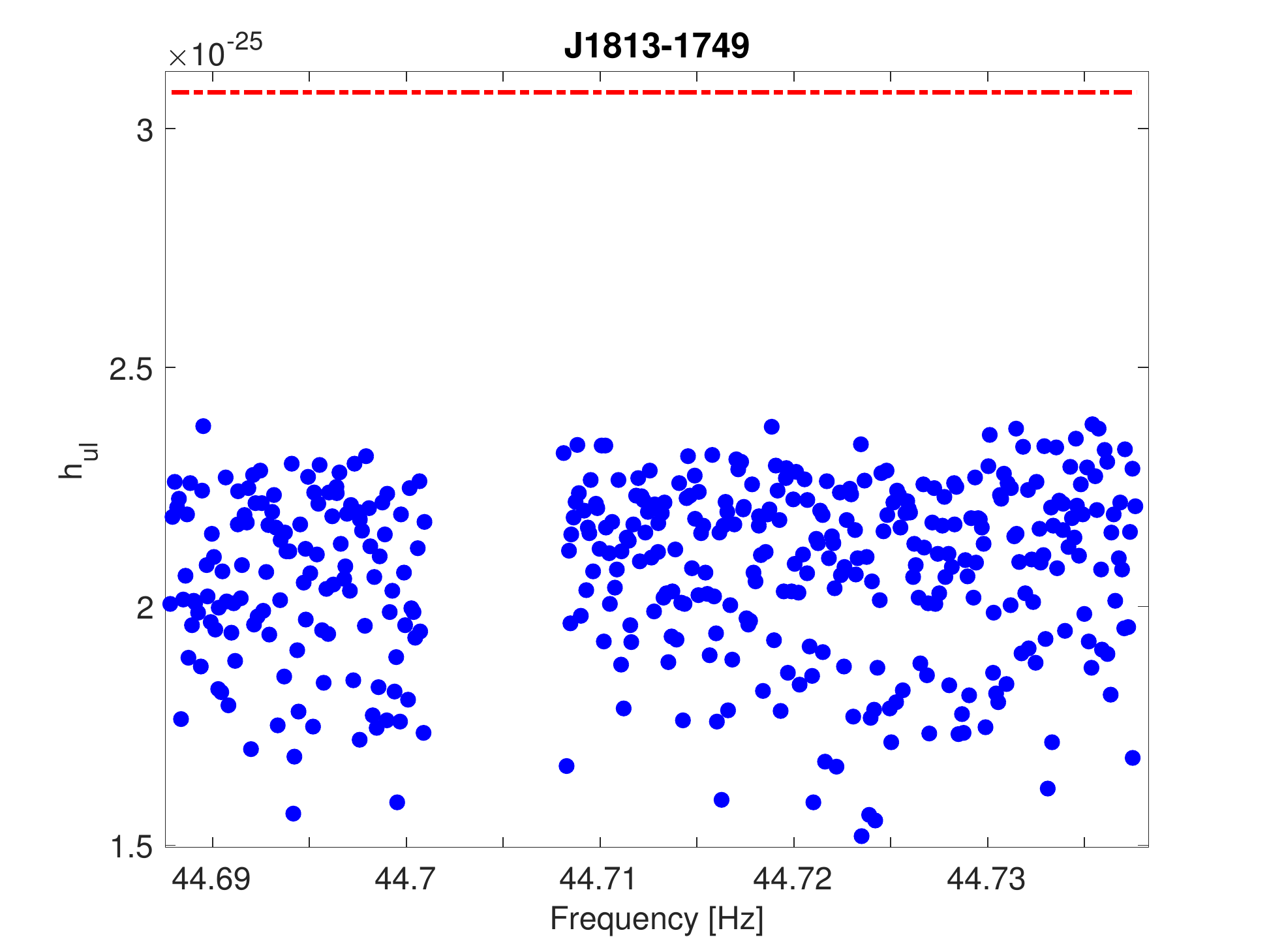}
\includegraphics[width=0.329\textwidth]{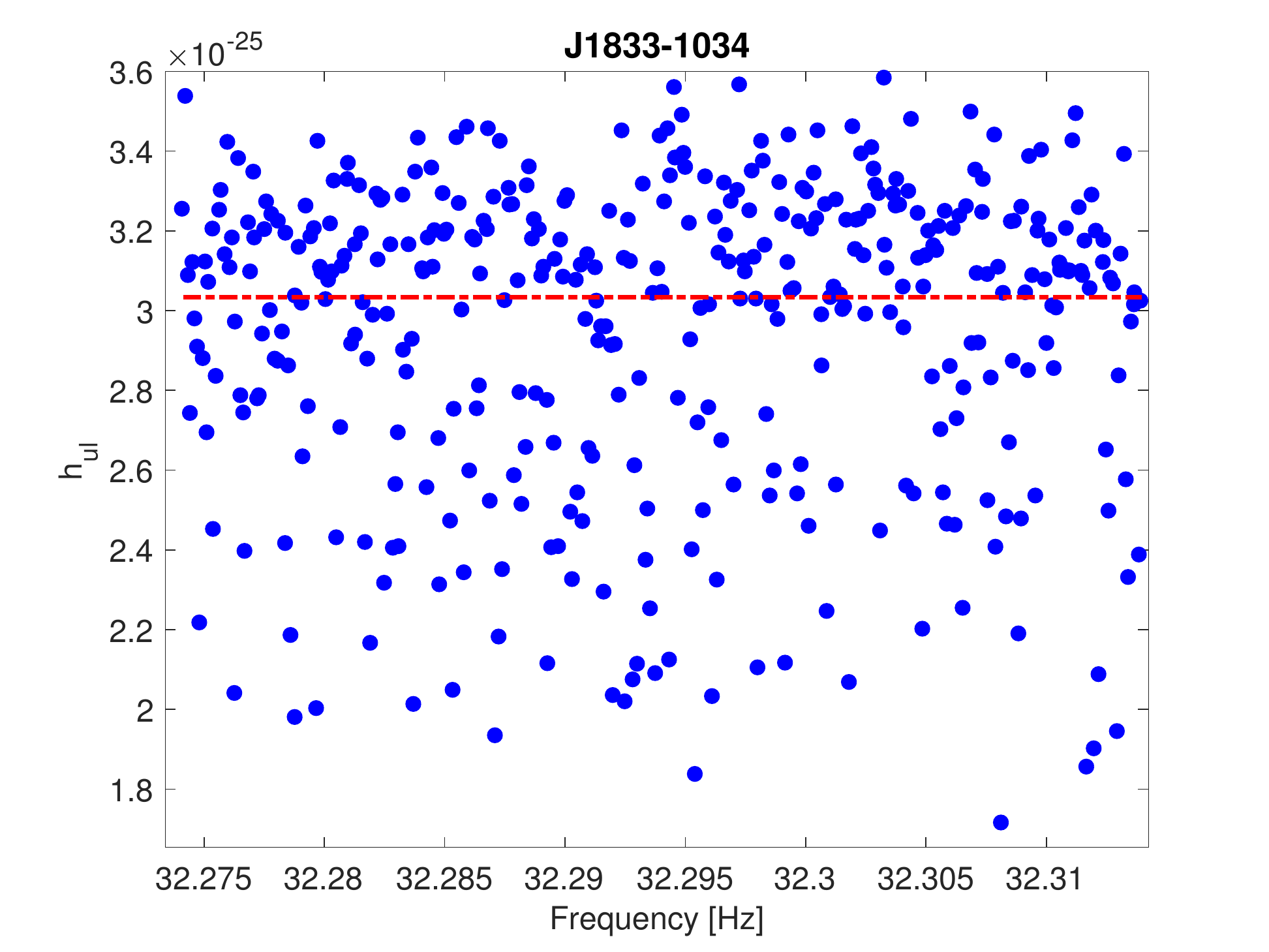}
\includegraphics[width=0.329\textwidth]{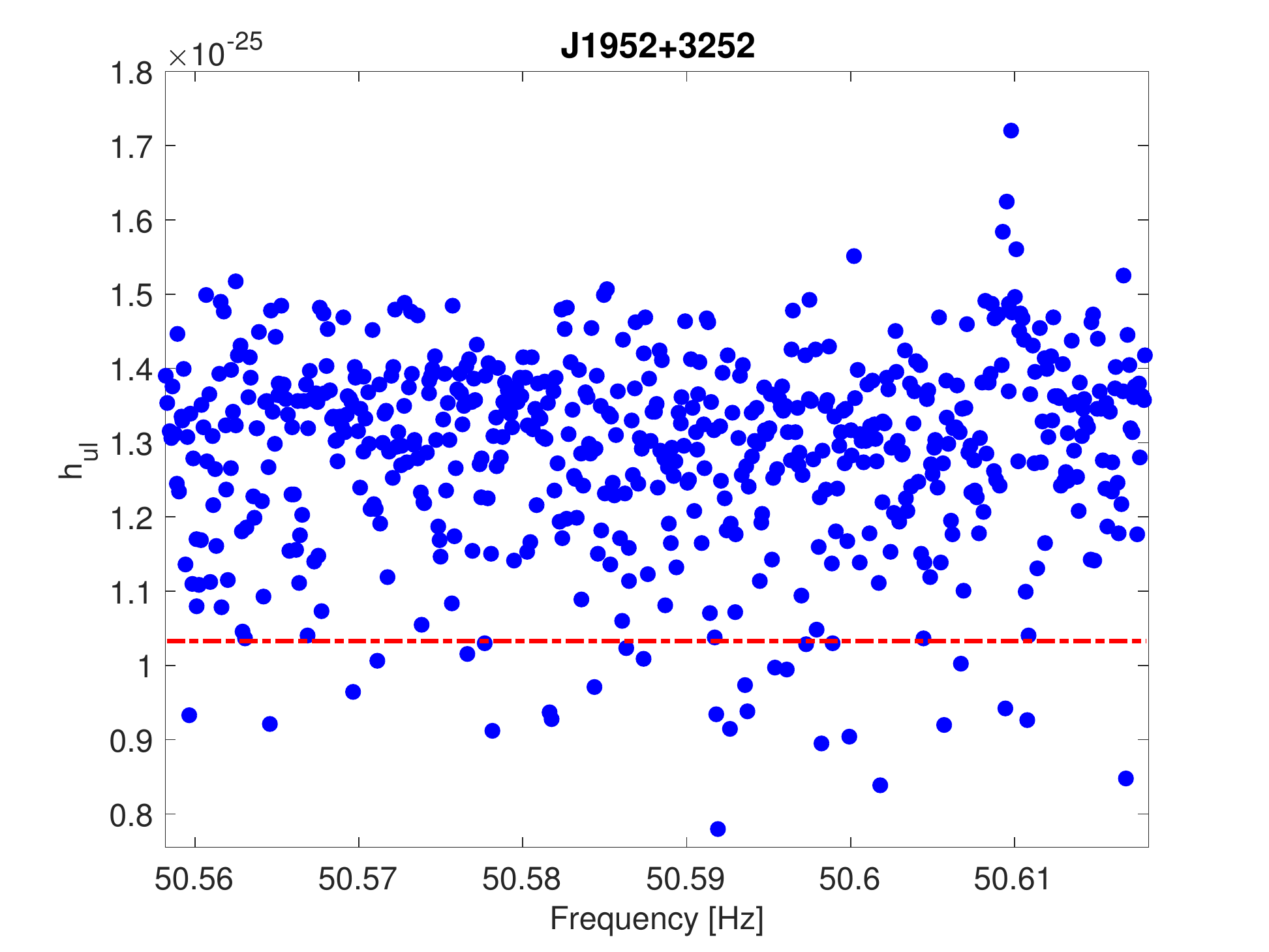}
\includegraphics[width=0.329\textwidth]{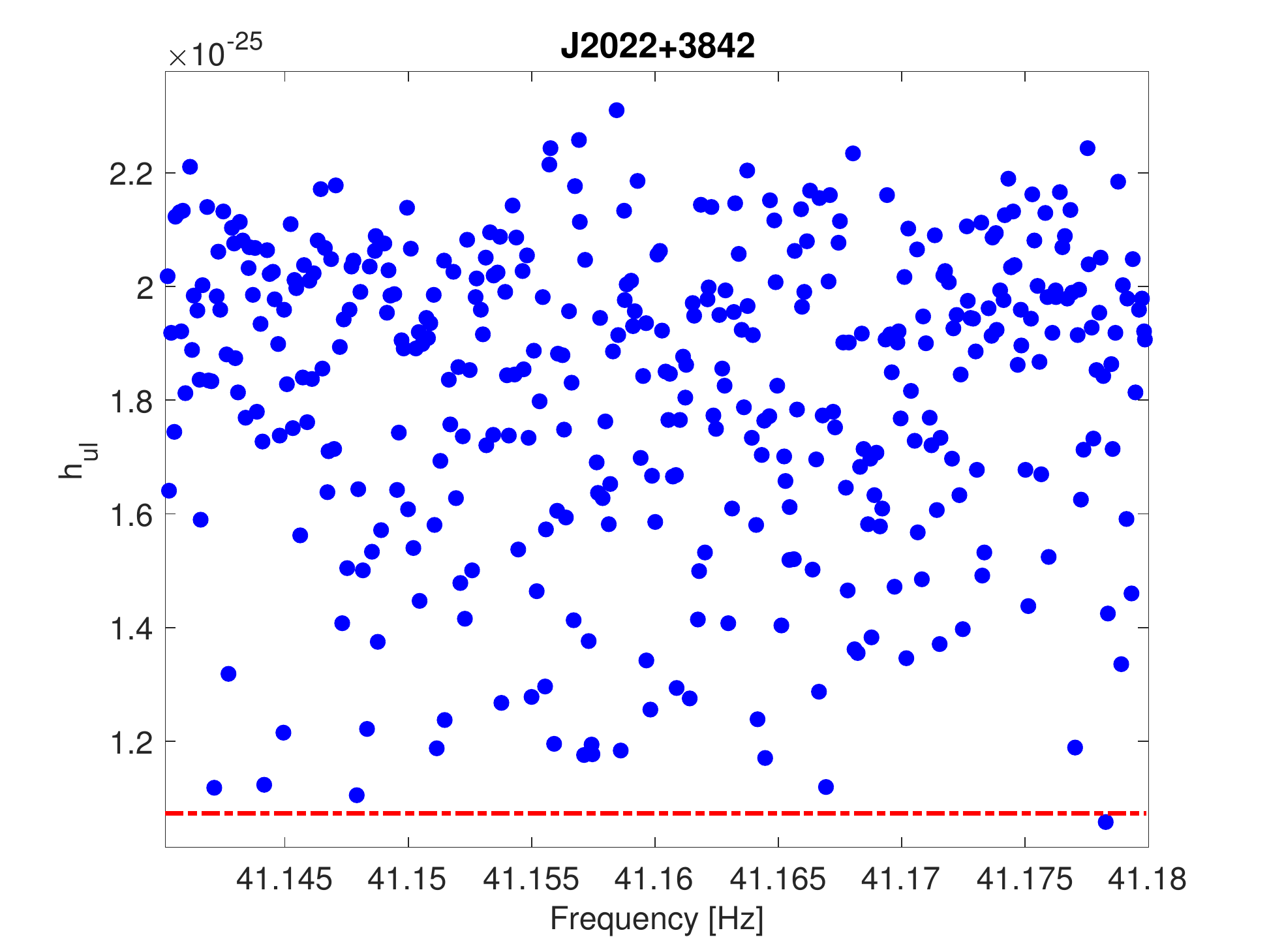}
\includegraphics[width=0.329\textwidth]{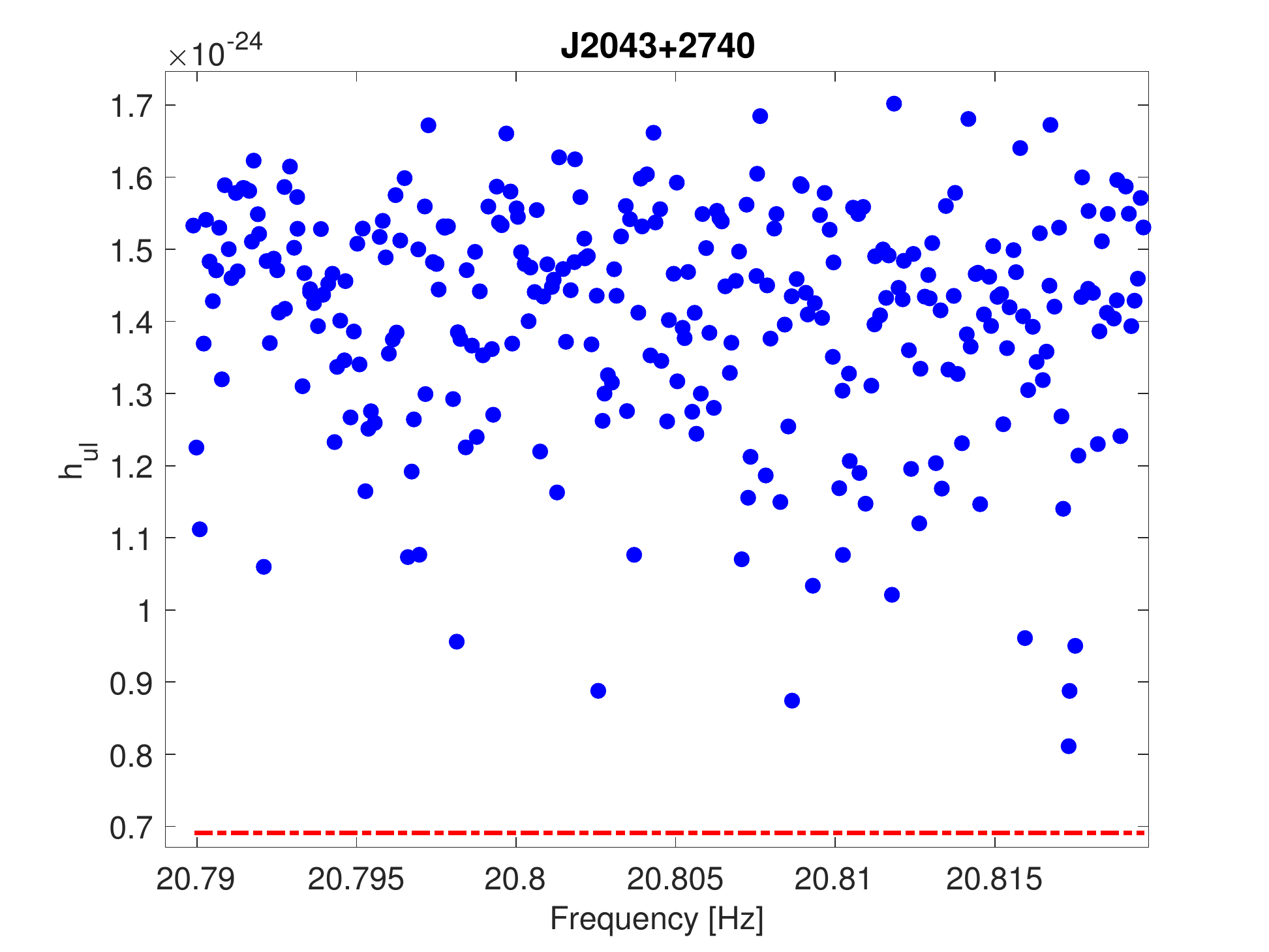}
\centering \includegraphics[width=0.329\textwidth]{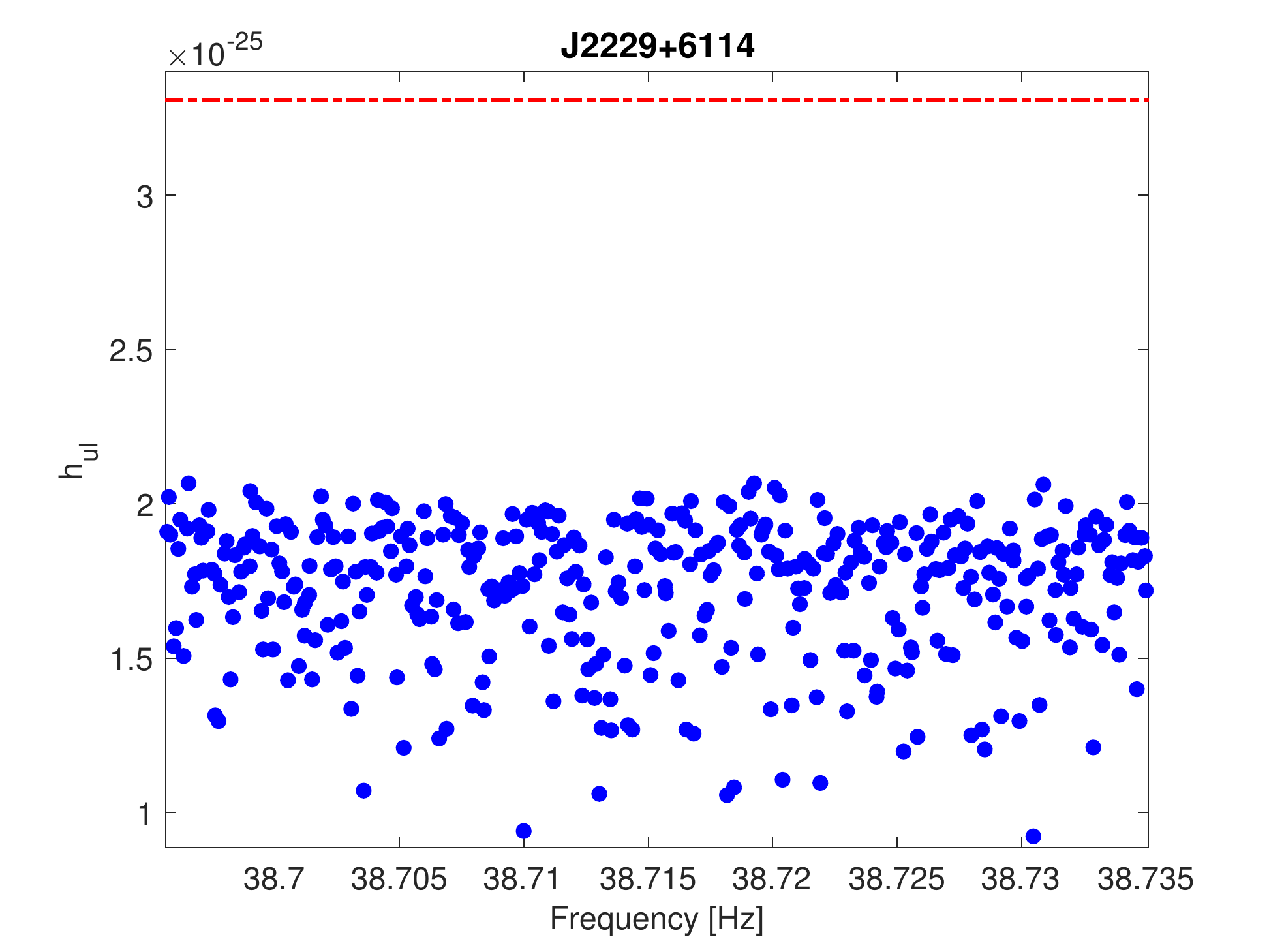}
\caption{Plots of the 95\% CL upper-limit on the GW ampitude for the 11 pulsars. The blue dots indicate the amplitude upper-limits set with our analysis, the red dashed line indicate the theoretical spin-down limit in Tab. \ref{tab:tab1}.}
\label{fig:datahist}
\end{figure}

\newpage
\appendix

\section{Upper-limit}
\label{app:upperl}
Once we have concluded that our data is compatible with noise, upper-limits on the GW amplitude can be computed. The upper-limits computation consists of injecting many different signals with fixed amplitude $H_0$ and parameters $\eta,\psi$ with a uniform distribution into the real data. According to the frequentist paradigm, the 95\% confidence level upper-limit can be computed asking that the 95\% of the injected signals  provide a value of the DS greater than the threshold for candidates selection used in the analysis. The signal must be injected at the beginning of the analysis, i.e. before the Doppler corrections and all the analysis procedure must be followed in order to compute the DS. This procedure is not suitable for narrow-band searches due to the fact that an injection is needed  in every analysed frequency sub-bands. This problem can be overcome by injecting simultaneously many different signals in many different frequency sub-bands in just one dataset and then perform the narrow-band search. Repeating this step $N$ times produce $N$ different datasets, each containing a signal in each analysed frequency sub-band. Then for each sub-band we ask for the 95\% DSs produced by the injected signal to be greater than the value used for the candidates selection, obtaining in this way the value of the upper-limit for a given frequency sub-band. Practically this procedure is done using several tricks in order to speed up the computation, as detailed in the following. First of all we assume that our data is the linear superposition of noise $n(t)$ and an injected signal $h_{\text{inj}}(t)$, namely $s(t)=n(t)+h_{\text{inj}}(t)
$. According to the linearity of the the FFT, the 5-vector of $s(t)$ will be the summation of the the two independent 5-vector of the noise and the injected signal:
\begin{equation}
\vec{X}=\vec{X}_{\text{noise}}+\vec{X}_{\text{inj}}\,.
\label{eq:FFT_linear}
\end{equation} The estimators of the GW polarisation, which are the building blocks of the DS, are linear due to the scalar product with respect to the sidereal templates $A_+(t),A_{\times} (t)$. Hence using Eq. \ref{eq:FFT_linear} we can write the analysis estimator as: 
  
\begin{equation}
\widehat{H}^{+ / \times}=\widehat{H}^{+ / \times}_{\text{noise}}+\widehat{H}^{+ / \times}_{\text{inj}}\, .
\label{eq:esti_linear}
\end{equation}Eq. \ref{eq:esti_linear} indicates that before the calculation of the DS we can keep separate the estimators computed from our real dataset and the ones arising from an injected signal. This leads to the possibility to change the GW amplitude $H_0$ of the injected signal directly re-scaling the absolute value of the estimators $\widehat{H}^{+ / \times}_{\text{inj}}$ without re-performing all the corrections in time domain  and thus saving computational time. As stressed before the form of the injected signal $h_{\text{inj}}(t)$ should be built in such a way to contain a signal in each analysed frequency sub-band. Formally we can write $h_{\text{inj}}(t)$ as the superposition of $N$ different signals each of one located in a random-frequency bin of each frequency sub-band.

\begin{equation}
h_{\text{inj}}(t)=H_0 [H^+ A_+ (t) + H^\times A_\times (t)] e^{i \phi_0} \sum_{\mathcal{S}=1}^{N} e^{i \phi^{\mathcal{S}}_{\text{R\"{o}m}}(t)} e^{i \phi^{\mathcal{S}}_{\text{rot}}(t)} \, ,
\label{eq:inj_full}
\end{equation}
where $\phi^{\mathcal{S}}_{\text{R\"{o}m}}(t)$ and $\phi^{\mathcal{S}}_{\text{rot}}(t)$ are the usual phase evolution due to the R\"{o}mer and rotational frequency evolution of the signal $\mathcal{S}$ \cite{rob:method}. Assuming that the $N$ different signals are injected with a constant frequency step $ \Delta f_{\text{inj}} $ in the frequency grid starting from a frequency $f_0$, i.e. $ f_{\mathcal{S}}=f_{0}+\mathcal{S} \Delta f_{\text{inj}} $, we can manipulate the Eq. \ref{eq:inj_full} to obtain:

\begin{eqnarray}
 h_{\text{inj}}(t)=H_0 [H^+ A_+ (t) + H^\times A_\times (t) ] e^{i \phi^{0}_{\text{R\"{o}m}}(t)}   e^{i \phi^{0}_{\text{rot}}(t)} e^{i \phi_0} \sum_{\mathcal{S}=1}^{N} e^{i 2 \pi \mathcal{S} \Delta f_{\text{inj} }(t+p(t))} \, ,
\label{eq:inj_redux}
\end{eqnarray}where $p(t)$ is the R\"{o}mer correction and the superscript $"0"$ refers to the phase evolution of a signal injected at the frequency $f_0$. By defining $k=2 \pi i \Delta f_{\text{inj}} (t+p(t)) $, we can now exploit the geometrical series present in Eq. \ref{eq:inj_redux} to write

\begin{eqnarray}
h_{\text{inj}}(t)=H_0 [H^+ A_+ (t) + H^\times A_\times (t) ] e^{i \phi^{0}_{\text{R\"{o}m}}(t)} e^{i \phi^{0}_{\text{rot}}(t)}  \dfrac{1-e^{ (N+1)k}}{1-e^k} 
\label{eq:inj_comp}
\end{eqnarray}Practically in our analysis, for each dataset, we select a random frequency bin in the first analysed frequency sub-band and then we replicate it on the frequency grid using Eq. \ref{eq:inj_comp} and setting $\Delta f_{\text{inj}}$ equal to the width of the sub-bands. This procedure together with the linearity of the FFT allow us to strongly reduce the computational time obtaining the same results.

\section{Known instrumental noise lines}

\begin{figure}[H]
\includegraphics[width=0.5\textwidth]{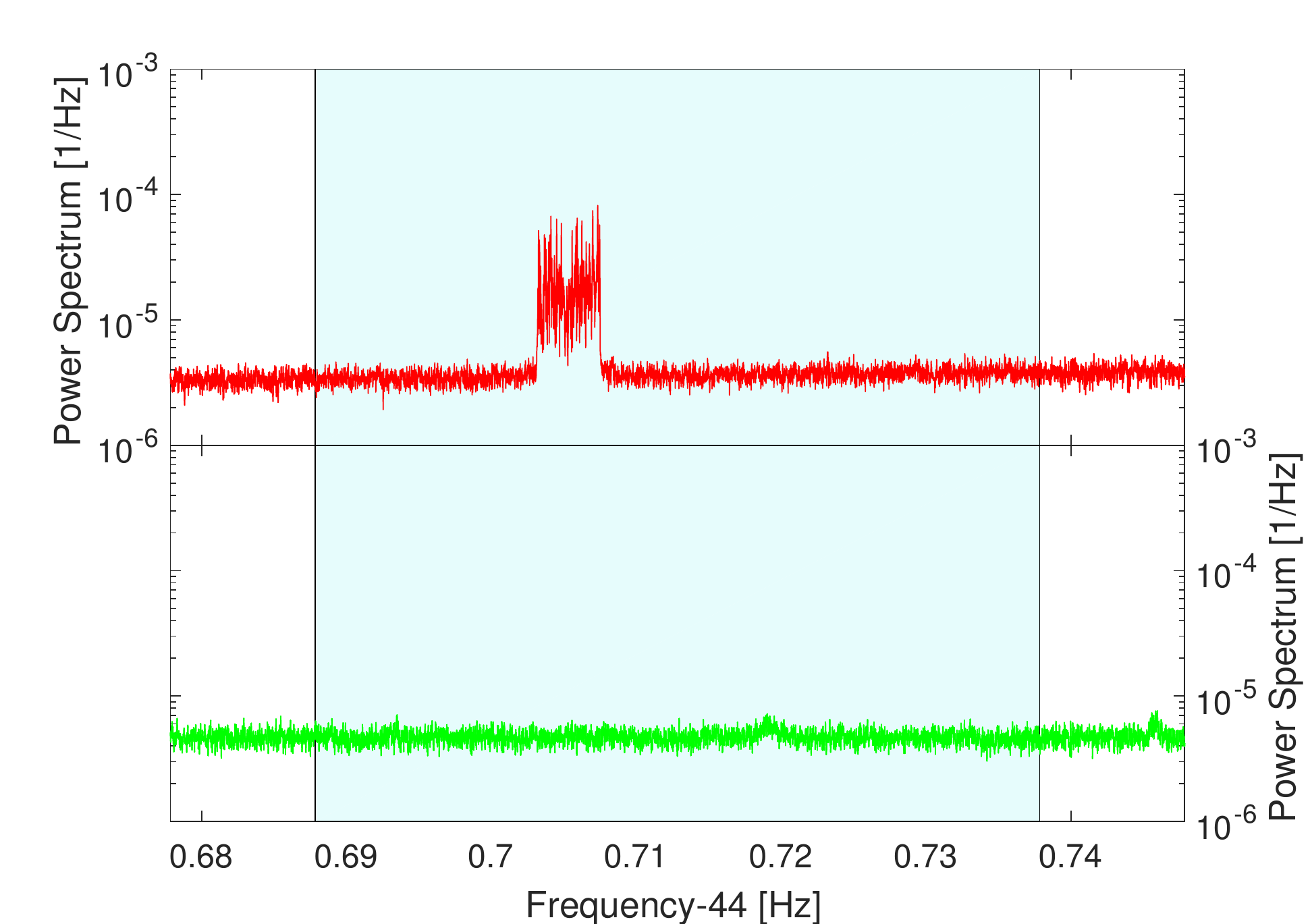}
\includegraphics[width=0.5\textwidth]{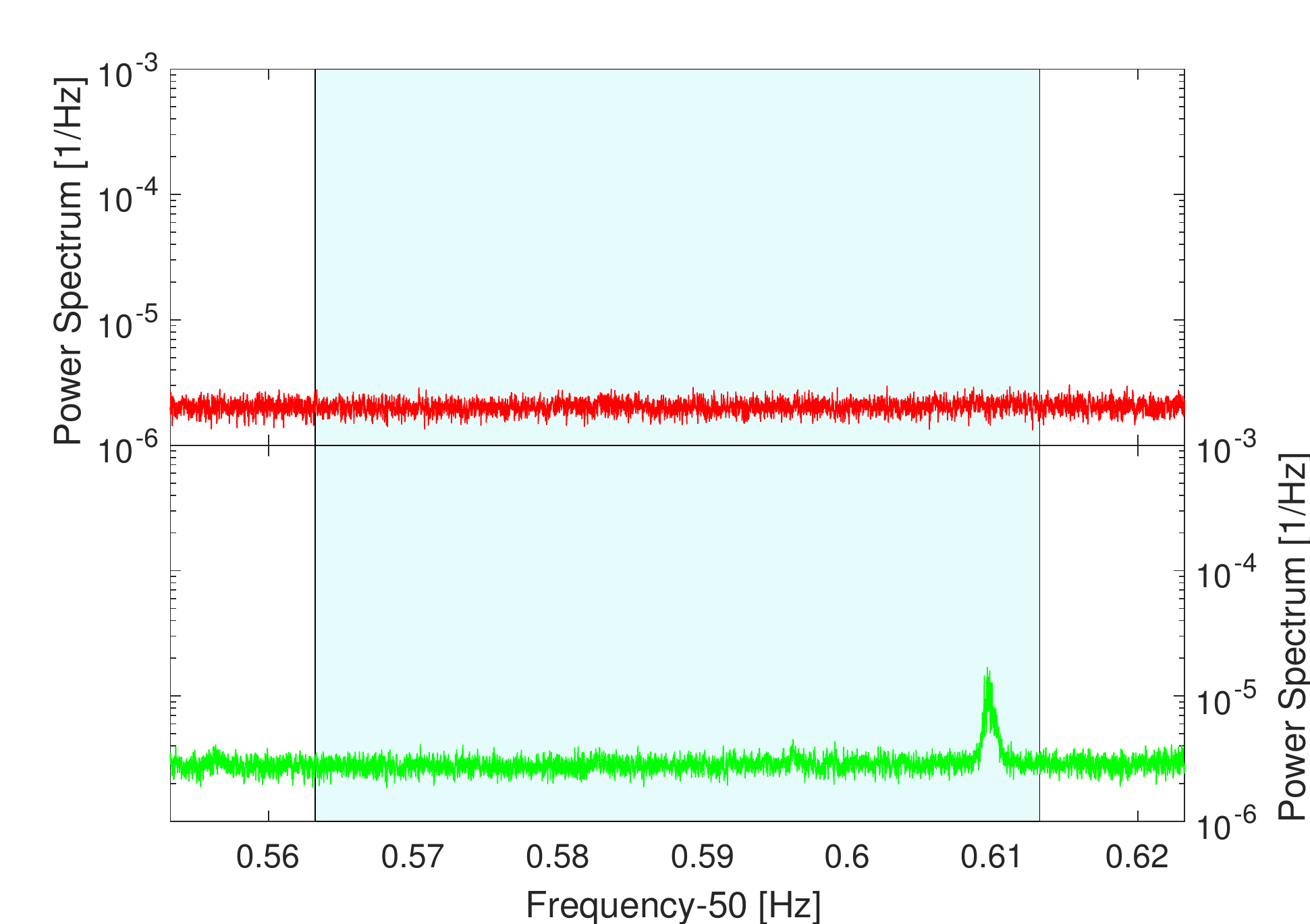}
\caption{Left: Power spectrum of Hanford (red line) and Linvingston (green line) data inside the frequency region explored by the narrow-band search (blue box) around J1813-1749. right: Power spectrum of Hanford (red line) and Linvingston (green line) data inside the frequency region explored by the narrow-band search (blue box) around J1953+3252.}
\label{fig:ps_O1_J1952}
\end{figure}

The data from the gravitational waves interferometer is polluted by several instrumental noise lines. Many of these disturbances have been identified during the run. Their presence can produce in the analysis a large number of outliers. We have found that the 36 outliers J1813-1749 are due to a noise line associated with the magnetometer channels in Hanford at $44.7029$ Hz. The presence of the noise line can also be seen in the left panel of  in Fig. \ref{fig:ps_O1_J1952}, where left plot where we show the power spectrum around the region explored by the narrow-band search. Concerning the 6 outliers from J1952+3252, we know that they are due to an artefact that is part of a $1.9464$ Hz comb in Livingston data. This disturbance is shown in the power spectrum in Fig.  \ref{fig:ps_O1_J1952} right plot.
  
\section{O2 follow-up of the outliers}

We have used these data in a narrow-band search in order to check if the outliers found for J1833-1034 and Vela in O1 were still present. The parameters of the narrow-band searches have been set in such a way to cover the expected frequency and spin-down of the outlier during the O2 epoch. The Vela pulsar glitched on  Dec 12th 2016 between 11:31 and 11:46 UT \footnote{http://www.astronomerstelegram.org/?read=9847}. The glitch have been classified as a canonical Vela-glitch \cite{hobbs:vela_glitch}. In order to prevent the glitch from affecting  our analysis we have started to analyse data from Jan 12th 2017 when the spin-down variation is supposed to be recovered. Moreover we have also increased  the spin-down range by a factor 3.7 with respect to the O1 analysis. A summary of the narrow-band search parameters is given in Tab. \ref{tab:O2}.
\setcounter{table}{6}

\begin{table}[H]
\caption{\label{tab:O2} This table reports the explored range for the rotational parameters of each pulsar. The columns are: the central frequency of the search ($f_0$), explored frequency band ($\Delta f$), central spin-down value of the search ($\dot{f}_0$), explored spin-down band ($\Delta \dot{f}_0$),the frequency ($f_{\text{O2}}$) and spin-down ($\dot{f}_{\text{O2}}$) of the outliers at the O2 epoch reference time on November 30th 2016.}
\begin{ruledtabular}
\begin{tabular}{cccccccc}
\textbf{Name}&$\mathbf{f_0}$ [Hz]&$\mathbf{\Delta f}$ [Hz]&$\mathbf{\dot{f}_0}$ [Hz/s]&$\mathbf{\Delta \dot{f}}$ [Hz/s] &$\mathbf{f_{\text{O2}}}$ [Hz]  &$\mathbf{\dot{f}_{\text{O2}}}$ [Hz/s] \\ \hline
\\
J0835-4510 (Vela) &$22.37289950$& $0.05$& $-3.1159 \cdot 10^{-11}$& $2.4024\cdot 10^{-13}$& $22.38712428$& $-3.1128 \cdot 10^{-11} $\\
J1833-1034 &$32.29004216$& $0.05$& $-1.0542 \cdot 10^{-10}$& $1.7266 \cdot 10^{-13}$& $32.27625775$& $-1.0534 \cdot 10^{-10}$ \\
\end{tabular}
\end{ruledtabular}
\end{table}
\vspace*{-10mm}
\begin{figure}[H]
\centering
\includegraphics[width=0.48\textwidth]{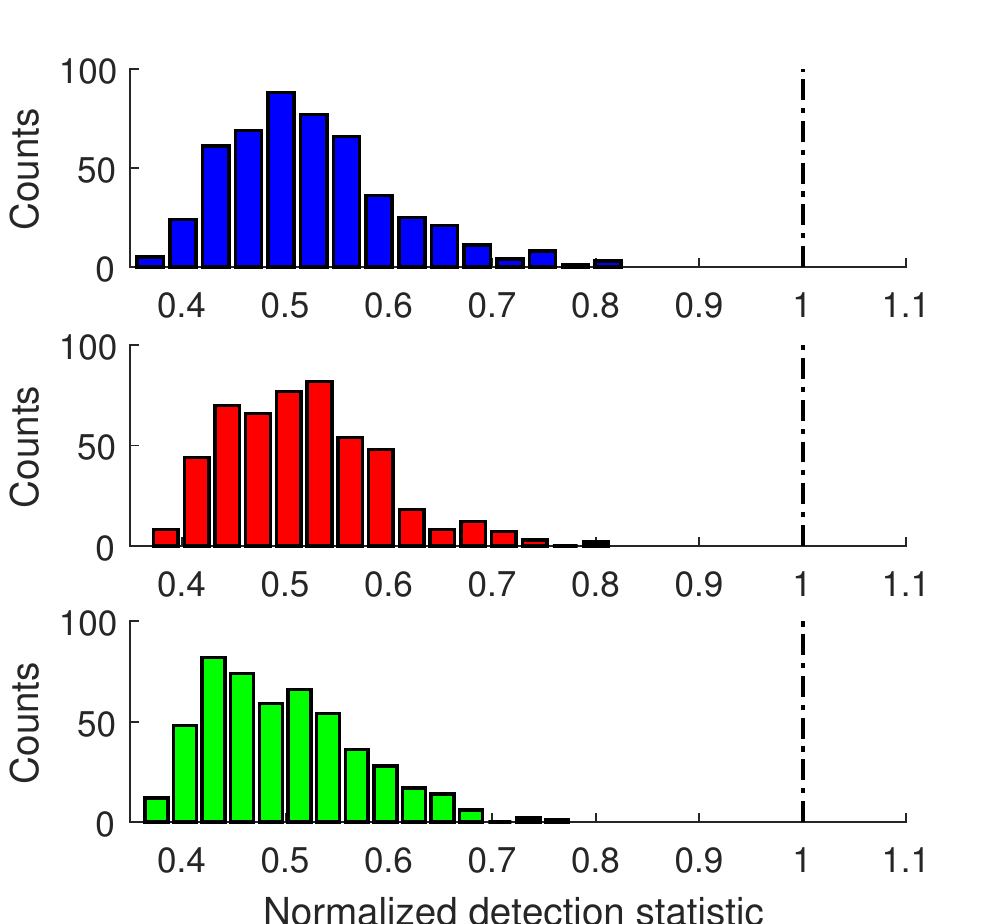}
\includegraphics[width=0.48\textwidth]{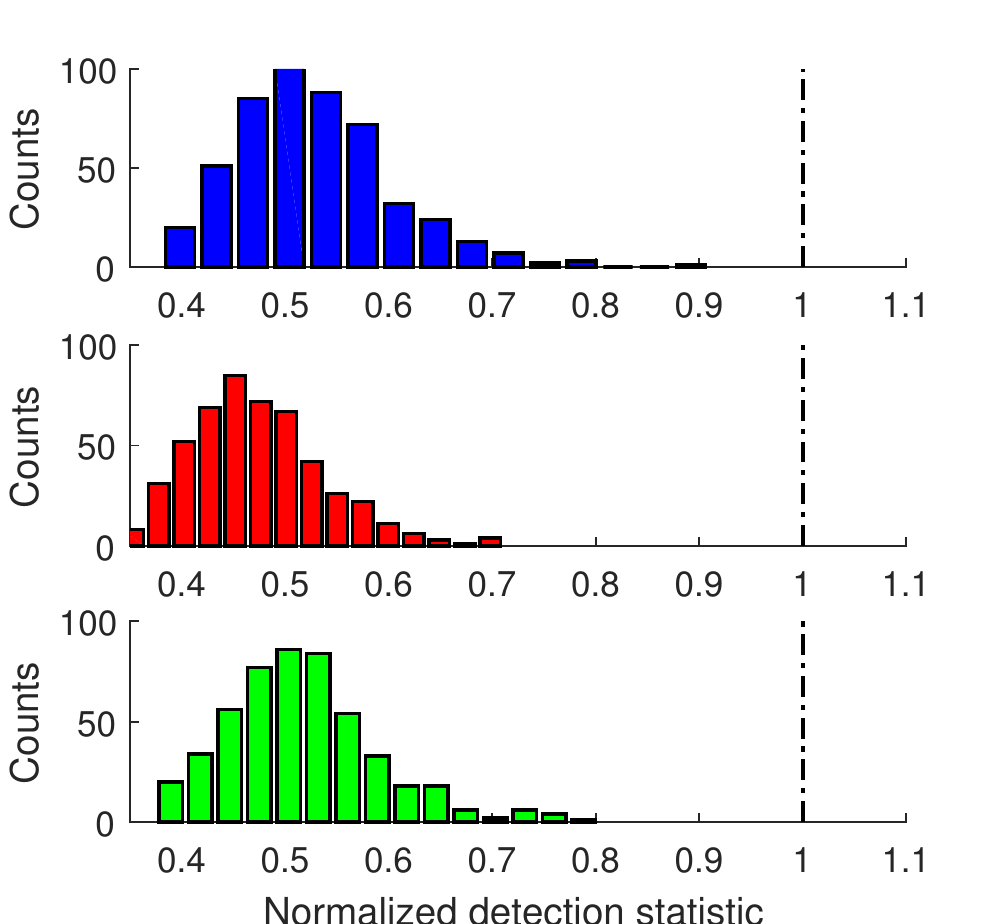}
\caption{Left: Histrograms of the DS obtained in J1833-1034 O2 narrow-band search, the x-axis is normalised to the DS threshold in each search. Top panel: Joint search, Middle panel: Hanford search, Bottom panel: Livingston search. Right: Histrograms of the DS obtained in Vela O2 narrow-band search, the x-axis is normalised to the DS threshold in each search. Top panel: Joint search, Middle panel: Hanford search, Bottom panel: Livingston search.}
\label{fig:Vela_O2}
\end{figure}

\begin{figure}[H]
\centering
\includegraphics[width=0.45\textwidth]{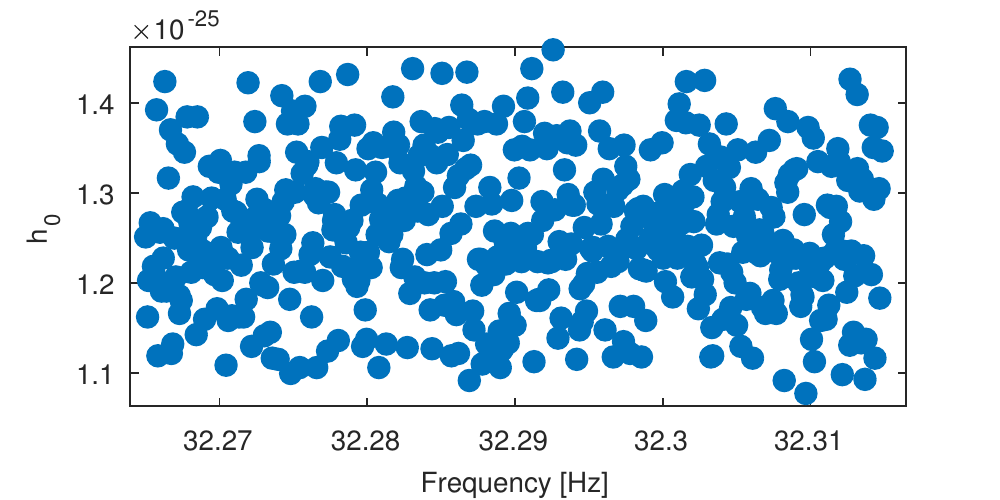}
\includegraphics[width=0.45\textwidth]{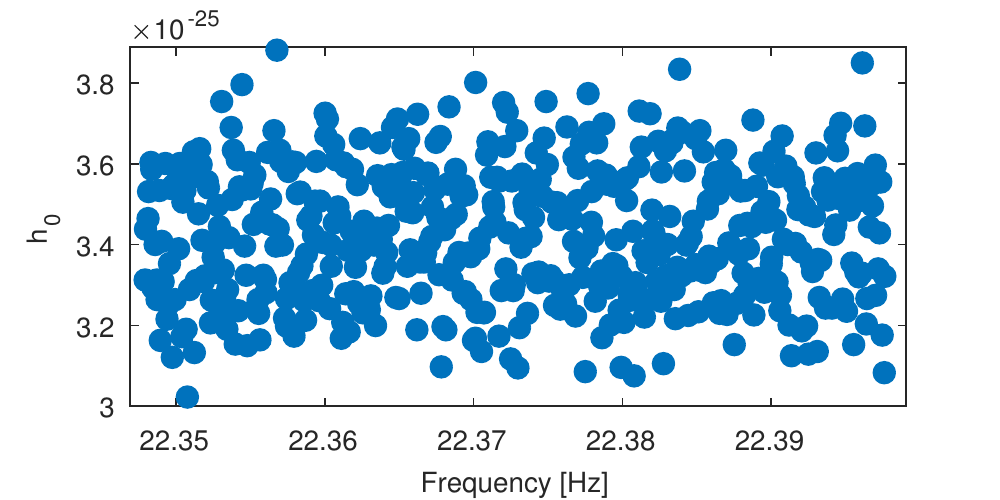}
\caption{Left: Upper-limits on the GW amplitude $h_0$ over the frequency narrow-region analysed in O2 for J1833-1034. Right: Upper-limits on the GW amplitude $h_0$ over the frequency narrow-region analysed in O2 for Vela}
\label{fig:Vela_O2_UL}
\end{figure}

Our analysis has produced no significant outlier for either J1833-1034 or Vela. Fig. \ref{fig:Vela_O2} shows the histograms of the DS obtained in the narrow-band search with respect to the threshold for outliers selection, for J1833-1034 and Vela respectively. In order to estimate our sensitivity in this search and compare the results with the sensitivity reached in O1, we have also computed the upper-limit on the GW amplitude $h_0$ for J1833-1034 and Vela over the narrow-frequency region explored. The procedure that we have used is the same used for O2, and the values of the upper-limits are shown in Fig. \ref{fig:Vela_O2_UL} for J1833-1034 and Vela respectively. The median value of the amplitude upper-limit for J1833-1034 is $1.25 \cdot 10^{-25}$ which is nearly a factor 2 lower than the one obtained for O1 analysis in Tab. \ref{tab:tab3}, thus indicating that if the outlier found in O1 were a true persistent CW signal,was a real it would have appeared in O2 analysis with an higher significance. Similarly, for Vela we have obtained a median value of the amplitude upper-limit of $3.41 \cdot 10^{-25}$ which is  about 3 times better than the one obtained in O1 analysis, see Tab \ref{tab:tab3}. We then conclude that both outliers are not confirmed in O2.

\end{document}